\documentclass[11pt]{article}
\usepackage{authblk}

\usepackage[T1]{fontenc}
\usepackage[utf8]{inputenc}
\usepackage{lmodern}
\usepackage{microtype}
\makeatletter
\providecommand\MT@suspend@tagging{}
\providecommand\MT@resume@tagging{}
\makeatother
\usepackage[a4paper,margin=1in]{geometry}
\usepackage{amsmath,amssymb,amsthm,mathtools}
\usepackage{booktabs,array}
\usepackage{enumitem}
\usepackage{xcolor}
\usepackage[compat=1.1.0]{tikz-feynman}
\usepackage{subcaption}
\usepackage{graphicx}

\usetikzlibrary{arrows.meta,calc,positioning,fit,decorations.pathreplacing}
\usepackage{hyperref}
\usepackage[nameinlink,capitalise]{cleveref}

\hypersetup{
  colorlinks=true,
  linkcolor=blue!55!black,
  citecolor=blue!55!black,
  urlcolor=blue!55!black,
  pdftitle={Working notes on the momentum Lagrangian amplituhedron}
}

\setlist[itemize]{topsep=3pt,itemsep=2pt,parsep=1pt}
\setlist[enumerate]{topsep=3pt,itemsep=2pt,parsep=1pt}
\allowdisplaybreaks

\newtheorem{definition}{Definition}[section]

\newtheorem{conjecture}[definition]{Conjecture}
\newtheorem{remark}{Remark}[section]

\newcommand{\Gr}{\operatorname{Gr}}
\newcommand{\GL}{\operatorname{GL}}
\newcommand{\LG}{\operatorname{LG}}
\newcommand{\Mat}{\operatorname{Mat}}

\newcommand{\pre}{\operatorname{pre}}
\newcommand{\inc}{\operatorname{inc}}

\newcommand{\cyc}{\mathrm{cyc}}

\newcommand{\Ampl}{\mathcal A}

\providecommand{\refl}{\mathrm{refl}}
\providecommand{\rot}{\mathrm{rot}}
\providecommand{\ip}{\mathsf{IP}}
\newcommand{\Lampl}{\mathcal M}

\newcommand{\st}{\mathord{*}}

\tikzset{
  blob/.style={draw,rounded corners=7pt,minimum height=1.2cm,minimum width=3.0cm,
               align=center,thick,fill=blue!4},
  core/.style={draw,rounded corners=7pt,minimum height=1.5cm,minimum width=3.4cm,
               align=center,thick,fill=green!5},
  tinyvertex/.style={circle,draw,inner sep=1.5pt,minimum size=5pt},
  internal/.style={thick,-{Latex[length=2mm]}},
  plaininternal/.style={thick},
  reflection/.style={densely dashed,gray,thick}
}
\title{{The Lagrangian Amplituhedron}}
\author{Syl Tom Krovi\thanks{Department of Mathematics, Weizmann Institute of Science, Rehovot, Israel. \texttt{syl.krovi@weizmann.ac.il} },  Ran J. Tessler\thanks{Department of Mathematics, Weizmann Institute of Science, Rehovot, Israel. \texttt{ran.tessler@weizmann.ac.il}}}

\begin{document}
\maketitle

\begin{abstract}
The article constructs the Lagrangian analog of the amplituhedron, studies its geometric properties and BCFW decompositions.
\end{abstract}
\tableofcontents
\section{Introduction}
\label{sec:introduction}
The amplituhedron was introduced by Arkani-Hamed and Tanka as a geometric
object whose canonical form encodes tree amplitudes in planar
\(\mathcal N=4\) super Yang--Mills theory
\cite{ArkaniHamedTrnka2014Amplituhedron}.  It replaces the usual
field-theoretic perspective on scattering amplitudes by a geometric
one - the amplitude as a generalized volume of a geometric space.  This point of view makes several structures which are hidden in
Feynman diagrams manifest, most notably Yangian invariance, recursive
BCFW decompositions, and sign-flip descriptions
\cite{BrittoCachazoFeng2005NewRecursion,BrittoCachazoFengWitten2005DirectProof,
ArkaniHamedBourjailyCachazoGoncharovPostnikovTrnka2016Grassmannian,
ArkaniHamedThomasTrnka2018Unwinding}.  More broadly, the amplituhedron became
one of the motivating examples for the theory of positive geometries and
canonical forms \cite{ArkaniHamedBaiLam2017PositiveGeometries}.

Despite this success, only a small number of amplituhedron-like geometries are
known.  Besides the original tree amplituhedron, important examples include
the \(m=1\) and \(m=2\) amplituhedra and their sign/tiling descriptions
\cite{KarpWilliams2019M1Amplituhedron,BaoHe2019M2Amplituhedron,
ParisiShermanBennettWilliams2023M2Amplituhedron}, the momentum
amplituhedron for planar \(\mathcal N=4\) SYM
\cite{DamgaardFerroLukowskiParisi2019MomentumAmplituhedron}, and the
orthogonal or ABJM amplituhedron
\cite{HuangKojimaWenZhang2022OrthogonalMomentumAmplituhedron,
HeKuoZhang2022MomentumAmplituhedronTwistorString}.  These examples share a
common architecture: a positive domain space, positive external data, a
kinematic image, natural boundary functions, BCFW-type tilings, and intrinsic
sign-flip characterizations.  At the same time, the ABJM example already shows
that the domain need not be an ordinary positive Grassmannian: orthogonality of
the domain changes both the combinatorics of the cells and the physical
boundary structure \cite{HuangWen2014ABJMPositiveOrthogonalGrassmannian,
GangHuangKohLeeLipstein2011ABJMRecursion}.  It is therefore natural to ask
whether amplituhedra are isolated miracles, or rather the first examples of a
larger family.

In this paper we construct a new amplituhedron-like geometry, which we call
the (reflected) Lagrangian amplituhedron. Its domain is Karpman's
reflected nonnegative Lagrangian Grassmannian, defined using an anti-diagonal
symplectic form and \emph{reflection-symmetric} plabic graphs
\cite{KarpmanLG,KarpmanSu}. The upcoming sequel \cite{Rot} we will constructs another Lagrangian amplituhedron, modelled over Shevchenko's rotational
nonnegative Lagrangian Grassmannian \cite{ShevchenkoRotationalLG}, whose visible symmetry is a \emph{half-turn rotation}
rather than a reflection.  In both cases the
amplituhedron map is
\[
\Phi_W:C\longmapsto CW,
\qquad
C\in\LG_{\geq}^{\diamond}(n,2n),
\qquad
\diamond\in\{\refl,\rot\},
\]
where \(W\in\Mat(2n,n+2)\) is positive external data.  The resulting
kinematics is a folded four-dimensional spinor-helicity kinematics:
the symplectic isotropy of the Lagrangian domain becomes momentum
conservation, while reflection or half-turn symmetry relates opposite
particles.

Our main finding is that the Lagrangian amplituhedron geometry satisfy the same kinds of
properties which make the known amplituhedra useful.  We identify its
natural Mandelstam-type boundary functions, describe the maximal domain cells
mapping to their boundary loci, formulate BCFW tilings, and give the expected
canonical-form recursions.  We also formulate a strengthened positivity assumptions on the
external data under which the boundary functions have definite sign, and the amplituhedron has a sign-flip description in the
\(B\)-amplituhedron/kinematic-picture.  We conjecture that Lagrangian amplituhedra are
positive geometries whose canonical forms are computed by the aforementioned BCFW
recursion.  In forthcoming work \cite{MS}, the BCFW tilings and sign-flip descriptions
will be proved under the strong-positivity hypotheses described below.  

The Lagrangian amplituhedron exhibits several new features. The most striking one
is that ordinary momentum amplituhedra appear as boundary factors. The reflected Lagrangian amplituhedron has certain boundary strata which factorize into a Lagrangian component and an ordinary momentum amplituhedron component. 
Thus the momentum amplituhedron is not
only a predecessor of our construction; it is literally a building block in
the factorization of Lagrangian amplituhedron boundaries. The other strata decompose as a product of two Lagrangian amplituhedra. 


In forthcoming works we will study other aspects of the Lagrangian amplituhedra, showing that the analogy with the previously discovered amplituhedra is not only in the level of BCFW decompositions and kinematics. 

\paragraph{\textbf{Structure of the paper.}}
This paper is structured as follows. Section 2 provides background on amplituhedra and Karpman's nonnegative Lagrangian Grassmannian. Section 3 defines the Lagrangian amplituhedron and the induced symplectic kinematics.
Section 4 studies the geometry of the reflected Lagrangian Grassmannian.

\paragraph{\textbf{AI disclosure}}
The authors have used ChatGPT 5.5 to assist with writing the
paper and running experiments verifying the mathematical claims. The authors take full responsibility for its mathematical content. The results appearing in the paper were found solely by the authors.

\paragraph{\textbf{Acknowledgments}}
The authors would like to thank Michael Oren-Prelstein for many fruitful discussion, and comments on this manuscript.
R.T. was supported by the ISF (grant No. 1729/23).

\section{Background}
\label{sec:background}

In this section we recall the small amount of background on total positivity and amplituhedra which will be needed below. We keep the discussion at tree level, and use it mainly to fix notation and to place the {Lagrangian amplituhedron introduced in this paper} in the existing amplituhedron family.

\subsection{The nonnegative Grassmannian}

Let $\Gr(k,n)$ denote the Grassmannian of $k$-planes in $\mathbb R^n$. We represent a point by a full-rank $k\times n$ matrix $C$, modulo left multiplication by $\GL_k$. For every $k$-element subset
$I=\{i_1<\cdots<i_k\}\subset[n]$, we write
\[
\Delta_I(C):=\det(C_{i_1},\ldots,C_{i_k})
\]
for the corresponding Pl\"ucker coordinate. The nonnegative Grassmannian is
\[
\Gr_{\geq}(k,n)
=
\{\,C\in \Gr(k,n): \Delta_I(C)\geq 0
\text{ for all } I\in\binom{[n]}{k}\,\}.
\]
Its open positive part $\Gr_{>}(k,n)$ is defined by the strict inequalities
$\Delta_I(C)>0$ for all $I$.

The basic cells of $\Gr_{\geq}(k,n)$ are positroid cells. A positroid cell is specified by declaring which Pl\"ucker coordinates vanish and requiring all remaining Pl\"ucker coordinates to be strictly positive. Equivalently, positroid cells may be encoded by decorated permutations, Grassmann necklaces, or (equivalence classes of) reduced plabic graphs
\cite{Postnikov2006TotalPositivity,ArkaniHamedBourjailyCachazoGoncharovPostnikovTrnka2016Grassmannian,lam2015totally}. A reduced plabic graph gives positive coordinates on the corresponding cell, and in these coordinates the cell carries a canonical logarithmic form
\[
\omega=\bigwedge_a d\log x_a.
\]

\subsection{Various amplituhedra}
\subsubsection{The (tree) Amplituhedron}
Fix integers $k,m,n$ with $0\leq k\leq n$ and $m\geq1$. Let
\[
Z:\mathbb R^n\longrightarrow \mathbb R^{k+m}
\]
be a positive linear map, represented by an $n\times(k+m)$ matrix whose ordered maximal minors are positive:
\[
\langle Z_{i_1}\cdots Z_{i_{k+m}}\rangle>0,
\qquad
1\leq i_1<\cdots<i_{k+m}\leq n.
\]
The tree amplituhedron is the image
\[
\mathcal A_{n,k,m}(Z)
=
\{\,Y=CZ:\ C\in\Gr_{\geq}(k,n)\,\}
\subset \Gr(k,k+m).
\]
For $m=4$, this geometry was introduced by Arkani-Hamed and Trnka in the study of planar $\mathcal N=4$ super Yang--Mills theory
\cite{ArkaniHamedTrnka2014Amplituhedron}. The expectation is that $\mathcal A_{n,k,4}(Z)$ is a positive geometry in the sense of \cite{ArkaniHamedBaiLam2017PositiveGeometries}, whose \emph{canonical form}, which is a meromorphic top form which satisfies a certain BCFW recursion \cite{BrittoCachazoFeng2005NewRecursion,BrittoCachazoFengWitten2005DirectProof,ArkaniHamedBourjailyCachazoGoncharovPostnikovTrnka2016Grassmannian}, encodes the corresponding tree-level superamplitude. 

The definition above lives in an auxiliary $Y$-space. There is also a useful projected picture, closer to physical kinematics. If one fixes a reference plane $Y=Y^\ast$ and projects the external data $Z_i$ modulo $Y^\ast$, one obtains an $m$-dimensional configuration $z_i$. More intrinsically, the \emph{$B$-amplituhedron} $\mathcal{B}_{n,k,m}$ \cite{KarpWilliams2019M1Amplituhedron} is the image of the map $C\mapsto z=(z_1\cdots z_n)$ given by \[z=C^\perp \cap W,\]where $W$ is the column span of $Z.$ Below we will use the terms kinematic picture and $B-$picture interchangeably. Karp and Williams prove that the $B-$amplituhedron is naturally homeomorphic to the ordinary amplituhedron. Their proof shows in fact that the two spaces are diffeomorphic, and their argument extends to all known amplituhedra, and the {amplituhedron we present in this paper}, without a change. 
\cite{ArkaniHamedThomasTrnka2018Unwinding} conjectured that in the kinematic space the $2|m$ amplituhedron is precisely the subspace of $\Gr(m,W)$ given by elementary sign conditions on the brackets
\[\langle {i_1},{i_2},\ldots, i_m\rangle_z:=\det(z_{i_1}z_{i_2}\ldots z_{i_m}).\]
\[\langle {i_1},{i_1+1}.{i_2},{i_2+1}\ldots,{i_{m/2}},{i_{m/2}+1}\rangle_z \geq 0\]and the sequence \[\langle 1,2,\ldots,m-1,m \rangle,\langle 1,2,\ldots ,m-1,m+1\rangle,\ldots \langle 1,2,\ldots ,m-1,n\rangle\]
has a prescribed number of sign flips.

\cite{KarpWilliams2019M1Amplituhedron} show that under the correspondence $z\leftrightarrow Y$ the projective vector of brackets $(\langle {i_1},{i_2},\ldots, i_m\rangle_z)_{i_1,\ldots,i_m}$ equals the projective vector of determinants $(\det(YZ_{i_1}Z_{i_2}\ldots Z_{i_m}))_{i_1,\ldots,i_m},$ obtained by completing a $k\times (k+m)$ matrix representation of $Y$ to a square matrix by adding rows $Z_{i_1},\ldots,Z_{i_m}$ of $Z.$ 
We will usually use the notation $\langle {i_1},{i_2},\ldots, i_m\rangle$ to denote the different types of brackets, and add $z,Y,Z$ only if they are not clear from context.

These brackets are also useful for describing the (real) codimension $1$ boundaries of $\Ampl_{n,k,4}$ which are the zero loci of the four-brackets
\[
\langle i\,i{+}1\,j\,j{+}1\rangle,
\]
with the usual cyclic conventions, inside the amplituhedron.

\subsubsection{The momentum amplituhedron}

The momentum amplituhedron is a spinor-helicity version of the tree amplituhedron for planar $\mathcal N=4$ SYM
\cite{DamgaardFerroLukowskiParisi2019MomentumAmplituhedron}. We briefly recall its definition, since it is the closest predecessor of the constructions in this paper.

For the $\mathrm{N}^{k-2}\mathrm{MHV}$ sector, one introduces bosonized spinor-helicity data
\[
\widetilde\Lambda\in \Mat(k+2,n),
\qquad
\Lambda\in \Mat(n-k+2,n),
\]
where $\widetilde\Lambda$ and $\Lambda^\perp$ are taken to be positive. Given
\[
C\in\Gr_{\geq}(k,n),
\]
one defines
\[
\widetilde Y=C\,\widetilde\Lambda\in \Gr(k,k+2),
\qquad
Y=C^\perp\Lambda\in \Gr(n-k,n-k+2),
\]
where $C^\perp$ denotes the orthogonal complement of $C$ in $\mathbb R^n$. The momentum amplituhedron is the image of the map \[C\mapsto(\tilde{Y},Y),\]
and it is denoted by
\[
\mathcal M_{n,k}
\subset
\Gr(k,k+2)\times \Gr(n-k,n-k+2).
\]
The two auxiliary planes determine ordinary two-component spinors by projection:
\[
\lambda_i=(Y^\perp\Lambda)_i,
\qquad
\widetilde\lambda_i=(\widetilde Y^\perp\widetilde\Lambda)_i.
\]
Equivalently, the $B-$picture is
\[\lambda=C\cap\Lambda,\qquad\tilde\lambda=C^\perp\cap \tilde\Lambda,\] hence $
\lambda\subset C,
~
\widetilde\lambda\subset C^\perp.
$ and thus
\[
\sum_{i=1}^n \lambda_i\widetilde\lambda_i=\sum_{i=1}^np_i=0,
\]
where 
\[p_i=\lambda_i\widetilde\lambda_i^{\mathsf T}.
\]is the $i$th momentum,
is just the usual momentum conservation condition. The equivalence with the $B-$picture allows us to view the momentum amplituhedron directly in spinor-helicity variables. In what follows we will refer to this projected spinor-helicity realization as the $B$-picture: the basic output is a pair of spinor configurations $(\lambda,\widetilde\lambda)$, or equivalently a momentum-conserving configuration of massless four-dimensional momenta.

Following \cite{EvenZoharLakrecTessler2025BCFWTriangulation} we will refer to homogeneous rational functions in the kinematical space as \emph{functionaries}. By the correspondence between amplituhedra and their $B-$picture these can also be thought of as functions on the amplituhedron map's image, and we will use the same term to describe both types of functions.

The \emph{boundary functionaries}, that is the functionaries whose zero loci describe the (real) codimension $1$ boundaries of the amplituhedron in either of these pictures are the adjacent brackets
\[
\langle i\,i{+}1\rangle,
\qquad
[i\,i{+}1],
\]
together with the planar Mandelstam variables $s_I,$ where $I$ is a cyclic interval, given by
\[
s_I
=
\sum_{\substack{a<b\\ a,b\in I}}
\langle ab\rangle[ab].
\]
The physical meaning of $s_I=0$ is that the corresponding $p_I$ is massless. In the amplituhedron space, as in the case of $\Ampl_{n,k,m}$, these are lifted to functionaries
\[
S_I
=
\sum_{\substack{a<b\\ a,b\in I}}
\langle Yab\rangle[\widetilde Yab].
\]
The sign-flip characterization of the momentum amplituhedron says, in one parity convention, that the sequence of angle brackets has $k-2$ sign flips, while the sequence of square brackets has $k$ sign flips, together with positivity of the relevant planar Mandelstam variables.

\subsubsection{The orthogonal ABJM amplituhedron}

A second important predecessor is the ABJM amplituhedron, or orthogonal momentum amplituhedron
\cite{HuangKojimaWenZhang2022OrthogonalMomentumAmplituhedron,HeKuoZhang2022MomentumAmplituhedronTwistorString}. This is a three-dimensional analogue of the momentum amplituhedron, designed for tree amplitudes of ABJM theory \cite{AharonyBergmanJafferisMaldacena2008ABJM}. Its domain is the positive orthogonal Grassmannian \cite{HuangWen2014ABJMPositiveOrthogonalGrassmannian} rather than the ordinary positive Grassmannian. Again it is conjectured to be a positive geometry with a canonical form defined recursively via the ABJM version of the BCFW recursion \cite{GangHuangKohLeeLipstein2011ABJMRecursion}. Correspondingly, the output kinematics are three-dimensional spinor-helicity variables, where a momentum is represented by a symmetric product
\[
p_i^{\alpha\beta}=\lambda_i^\alpha\lambda_i^\beta.
\]
Momentum conservation arises from the orthogonality condition on the domain point.

Orthogonality imposes a self-duality constraint on the domain, and this changes both the combinatorics of the cells and the structure of the physical boundaries. In the ABJM case, the codimension-one physical boundaries are governed by odd-particle planar Mandelstam variables, while even-particle channels yield corners of higher codimension. 

\subsubsection{BCFW Tilings and sign flips}
The standard method for calculating the canonical form is to \emph{tile} the amplituhedron by images of positroid cells. Concretely,
we say that a collection of open subspaces $X_1,\ldots,X_N$ of a space $X$ tile $X$ if all pairwise intersections between $X_i\cap X_j,~i\neq j$ are empty, and $X$ is the union of closures of the different $X_i.$ We will also use the closely related terms of a tile and a tilings, in the context of amplituhedra. If a positroid cell in the domain Grassmannian of an amplituhedron map if of the same dimension as the target amplituhedron, and maps injectively to it, we will refer to its image as a \emph{tile}. A \emph{tiling} is a collection of such cells whose images tile the amplituhedron.
Suppose a collection of positroid cells
\[
\Pi_1,\ldots,\Pi_r\subset \Gr_{\geq}(k,n)
\]
gives a tiling. 
It is conjectured, e.g. in \cite{ArkaniHamedBaiLam2017PositiveGeometries}, which defines the notion of positive geometry, that their images are positive geometries, that the canonical form of the tile is the pushforward of the cell form, and, importantly, the canonical form of the corresponding amplituhedron is obtained by summing over tiles:
\[
\Omega(\mathcal A)=\sum_a (\Phi_Z)_*\omega_{\Pi_a}.
\]

The most important tilings are the BCFW tilings. They are geometrizations of the BCFW recursion for scattering amplitudes and their differential forms. For the $m=4$ tree amplituhedron, the standard BCFW conjecture was proved in for standard BCFW cells
\cite{EvenZoharLakrecTessler2025BCFWTriangulation}, and in
\cite{EvenZoharLakrecParisiShermanBennettTesslerWilliams2025BCFWTilings} for more general BCFW tilings. The case of the momentum amplituhedron, under appropriate strengthened positivity assumptions, was proved by Galashin \cite{Galashin2024AmplituhedraOrigami}. For the ABJM amplituhedron, the BCFW tiling conjecture was proved under the analogous positivity assumptions in
\cite{OrenPerlsteinTessler2025ABJMBCFWTiling}. One reason why stronger positivity assumptions are needed is that without them Mandelstam variables are not necessarily positive, and as a consequence it is harder to describe the boundaries in this setting.
The above papers also derived the aforementioned  boundary structure of the physical amplituhedra. The recent \cite{Galashin2026AmplituhedraOrigamiLoopLevel} proved the BCFW conjecture in the loop level for the ordinary and momentum amplituhedra, the latter under the strengthened positivity assumptions.

Sign-flip descriptions of amplituhedra provide intrinsic characterizations of this space. For the ordinary $m=4$ amplituhedron, Arkani-Hamed, Thomas and Trnka formulated a purely sign-theoretic description and proved one inclusion
\cite{ArkaniHamedThomasTrnka2018Unwinding}. The original momentum amplituhedron paper \cite{DamgaardFerroLukowskiParisi2019MomentumAmplituhedron} formulated the analogous statement in spinor-helicity variables
\cite{DamgaardFerroLukowskiParisi2019MomentumAmplituhedron}, and the original ABJM papers formulated the corresponding orthogonal version
\cite{HuangKojimaWenZhang2022OrthogonalMomentumAmplituhedron,HeKuoZhang2022MomentumAmplituhedronTwistorString}. In all cases the argument of \cite{ArkaniHamedThomasTrnka2018Unwinding} provided one side. For the $m=2$ amplituhedron the characterization was proven in \cite{ArkaniHamedThomasTrnka2018Unwinding,BaoHe2019M2Amplituhedron,ParisiShermanBennettWilliams2023M2Amplituhedron}. For the $m=4$ amplituhedron, and for its momentum-space variant, under the relevant strengthened positivity hypotheses, these descriptions have been proved by Galashin
\cite{Galashin2024AmplituhedraOrigami}. The upcoming \cite{AEZLPSBTW} shows however that the sign flip conjecture is generally false.

\subsection{Ordinary momentum-amplituhedron factorization}
\label{subsec:ordinary-momentum-factorization}
We finish the background by recalling the ordinary momentum-amplituhedron boundary
factorization pattern.  This construction will be instrumental in the study of
{Lagrangian amplituhedron} boundary decompositions.

\subsubsection{The factorization boundary}

Let the external labels be divided into two cyclic intervals
\[
I_L\sqcup I_R=[n].
\]
Consider the Mandelstam boundary
\[
s_{I_L}=0.
\]
At a generic point of this boundary, the partial momentum
\[
P_{I_L}:=\sum_{i\in I_L}p_i
\]
has rank one.  We introduce an internal massless particle \(\star\) with
\[
p_\star=-P_{I_L}=P_{I_R}.
\]
The two lower kinematic configurations live on
\[
I_L\cup\{\star\},
\qquad
\{\star\}\cup I_R,
\]
with the internal momentum taken with opposite orientation on the two sides.

The boundary component is described by two lower ordinary momentum
amplituhedra glued along this internal particle.  Choose a sector
\[
\mathcal M_{N_L,k_L}
\]
on the left, where
\(
N_L=|I_L|+1.
\)
The right factor has
\[
N_R=|I_R|+1,
\qquad
N_L+N_R=n+2.
\]
Its Grassmannian rank is forced by the requirement that the glued domain
plane have rank \(k\).  Indeed, if the right factor has rank \(k_R\), then
\emph{amalgamating}  the two domain spaces along the internal leg (see below) produces a space of
rank
\[
k_L+k_R-1.
\]
Thus, in order to obtain a point of \(\Gr_{\geq}(k,N)\), one must have
\[
k_L+k_R-1=k,
\]
or equivalently
\[
k_L+k_R=k+1.
\]
Thus the right sector is
\[
k_R=k+1-k_L.
\]
In this notation, the generic factorization boundary is built from
\[
\mathcal M_{N_L,k_L}
\qquad\text{and}\qquad
\mathcal M_{N_R,k_R},
\qquad
k_R=k+1-k_L,
\]
glued over the common internal on-shell particle.

\subsubsection{Positive-cell amalgamation}

At the level of positive Grassmannian cells, take positroid cells
\[
\Pi_L\subset\Gr_{\geq}(k_L,N_L),
\qquad
\Pi_R\subset\Gr_{\geq}(k_R,N_R),
\]
with boundary labels
\[
I_L\cup\{\star_L\},
\qquad
\{\star_R\}\cup I_R.
\]
Graphically, one takes reduced plabic graphs \(G_L\) and \(G_R\), identifies
the two internal boundary legs \(\star_L,\star_R\), and turns the identified
leg into an internal edge.  This is the positive-Grassmannian amalgamation
operation of
\cite{ArkaniHamedBourjailyCachazoGoncharovPostnikovTrnka2016Grassmannian}.
Algebraically, it is a direct product followed by a projection.

Let
\[
V_L\subset\mathbb R^{I_L\cup\{\star_L\}},
\qquad
V_R\subset\mathbb R^{\{\star_R\}\cup I_R}
\]
be the two lower row spaces.  Their amalgamation is the fiber product
\[
V_L\star V_R
=
\left\{
(x_L,x_R):
\exists t\in\mathbb R,\ 
(x_L,t)\in V_L,\ 
(t,x_R)\in V_R
\right\}.
\]
Equivalently, one takes
\[
(V_L\oplus V_R)\cap\{x_{\star_L}=x_{\star_R}\}
\]
and then deletes the two internal coordinates.  For generic points,
\[
\dim(V_L\star V_R)=k_L+k_R-1=k.
\]
In suitable gauges the two lower domain matrices may be written as
\[
C_L=
\begin{pmatrix}
L_0&0\\
\ell&1
\end{pmatrix},
\qquad
C_R=
\begin{pmatrix}
1&r\\
0&R_0
\end{pmatrix}.
\]
After amalgamation, the resulting \(k\)-plane is represented by
\[
C_{\mathrm{fac}}
=
\begin{pmatrix}
L_0&0\\
\ell&r\\
0&R_0
\end{pmatrix}.
\]
The row \((\ell,r)\) is the shared internal row.  It is the domain-space
shaddow of the internal on-shell particle.  Dually, if \(U_L\) and \(U_R\) are
the column spans supported on the two external label sets, then
\[
\dim U_L=k_L,
\qquad
\dim U_R=k_R,
\qquad
\dim(U_L\cap U_R)=1,
\]
and the common line is the internal particle.

If the two lower positroid cells have  dimensions \(d_L\) and \(d_R\),
then the amalgamated cell has dimension
\[
d_L+d_R-1.
\]
The subtraction by one comes from identifying the two copies of the internal
leg, or equivalently from quotienting the common internal little-group
rescaling.

In the BCFW situation, the lower cells are chosen to yield
tiles of the lower momentum amplituhedra.  Thus their domain dimensions agree
with the dimensions of the corresponding lower images:
\[
d_L=\dim\mathcal M_{N_L,k_L}=2N_L-4,
\qquad
d_R=\dim\mathcal M_{N_R,k_R}=2N_R-4.
\]
Since
\[
N_L+N_R=n+2,
\]
the amalgamated source cell has dimension
\[
(2N_L-4)+(2N_R-4)-1=2n-5,
\]which is of codimension $1$ in the image.
\subsubsection{A six-point example}

For \(n=6\), \(k=3\), and the channel \(s_{123}=0\), one of the standard
seven-dimensional positive cells appearing in the momentum-amplituhedron
factorization analysis is
\[
C=
\begin{pmatrix}
1&\alpha_5+\alpha_7&\alpha_5\alpha_6&0&0&0\\
0&1&\alpha_6&\alpha_2+\alpha_4&\alpha_2\alpha_3&0\\
0&0&0&1&\alpha_3&\alpha_1
\end{pmatrix},
\qquad
\alpha_i>0.
\]
It satisfies
\[
\Delta_{123}=\Delta_{456}=0.
\]
Thus the first three and last three columns each have rank two.  The first two
rows give the left \(k_L=2\) block, the last two rows give the right \(k_R=2\)
block, and the middle row is the overlap.  Inserting one internal column into
each block recovers two positive \(2\times4\) matrices, each describing a
four-particle momentum-amplituhedron factor.  The seven positive parameters
agree with the factorization dimension
\[
2\cdot6-5=7.
\]

\subsection{Karpman's nonnegative Lagrangian Grassmannian}

We now recall the version of the totally nonnegative Lagrangian Grassmannian introduced by Karpman.  This will be the domain of {the Lagrangian amplituhedron studied in this paper}.

{We use the decorations $\refl$, since in the sequel \cite{Rot} we will study the rotational Lagrangian model, based on Shevchenko's Lagrangian Grassmannin \cite{ShevchenkoRotationalLG}.} When only one model is under discussion, we will occasionally omit these decorations from the notation.

Let the standard basis of $\mathbb R^{2n}$ be labeled by
\[
[2n]=\{1,2,\ldots,2n\},
\]
and let
\[
\bar i:=2n+1-i
\]
be the reflection involution.  We use the anti-diagonal symplectic form
\[
(E_{\refl})_{ij}=(-1)^j\delta_{i,\bar j}.
\]
Equivalently,
\[
\langle e_i,e_j\rangle_{E_{\refl}}
=
\begin{cases}
(-1)^j,& j=\bar i,\\
0,& j\neq \bar i.
\end{cases}
\]
Thus
\[
E_{\refl}^{\mathsf T}=-E_{\refl},
\qquad
E_{\refl}^2=-I_{2n}.
\]
This is Karpman's convention; it is adapted to the standard type-$A$ positivity convention on the ambient Grassmannian.

A point $C\in \Gr(n,2n)$ is Lagrangian if
\[
C E_{\refl} C^{\mathsf T}=0,
\]
where $C$ denotes any full-rank $n\times 2n$ matrix representative of the row space.  We write
\[
\LG^{\refl}(n,2n)
=
\{\,C\in\Gr(n,2n): C E_{\refl} C^{\mathsf T}=0\,\}.
\]
The nonnegative part is defined by the embedding into the ordinary Grassmannian:
\[
\LG^{\refl}_{\geq}(n,2n)
=
\LG^{\refl}(n,2n)\cap \Gr_{\geq}(n,2n).
\]
Thus all ordered maximal minors of $C$ are required to be nonnegative.  The open positive part is
\[
\LG^{\refl}_{>}(n,2n)
=
\LG^{\refl}(n,2n)\cap \Gr_{>}(n,2n).
\]
The dimension is
\[
\dim \LG^{\refl}(n,2n)=\frac{n(n+1)}{2}.
\]

We will repeatedly use one simple linear-algebraic consequence of the convention above.  If $C\in\LG^{\refl}(n,2n)$, then, with respect to the ordinary Euclidean dot product,
\[
C^\perp=C E_{\refl}
\]
as row spaces.  In other words, the Euclidean orthogonal complement of a Lagrangian plane is obtained from the plane itself by applying the symplectic matrix $E_{\refl}$.  This identity is the basic reason that a single positive Lagrangian domain produces a folded pair of chiral and antichiral spinor data below.

\subsubsection{Symmetric plabic graphs and positive operations}\label{subsub:refl_positive_ops}
Karpman's cells in $\LG^{\refl}_{\geq}(n,2n)$ admit plabic graph charts which are compatible with the reflection symmetry.  A \emph{symmetric plabic} graph is a plabic graph with $2n$ boundary vertices and a distinguished reflection axis whose endpoints lie in the two gaps
\[
2n\mid 1,
\qquad
n\mid n+1.
\]
Reflection through this axis preserves the underlying uncolored graph and reverses the colors of all internal vertices.  A symmetric weighting assigns equal weights to reflected edges.  The boundary measurement of a symmetrically weighted graph lies in $\LG^{\refl}(n,2n)$, and positive symmetric weights give positive charts for cells of $\LG^{\refl}_{\geq}(n,2n)$ \cite{KarpmanLG,KarpmanSu}.

For us, the most important point is not only the existence of these charts, but the elementary operations from which they are built.  We will use two kinds of positivity-preserving operations: reflected loop-coloop insertions and reflected bridge operations.

First recall the ordinary positroid operations $\operatorname{inc}_i$ and $\operatorname{pre}_i$.  The operation $\operatorname{inc}_i$ inserts a coloop at the boundary label $i$, represented graphically by a white lollipop, while $\operatorname{pre}_i$ inserts a loop at $i$, represented by a black lollipop.  In the Lagrangian setting these operations must be inserted in reflected pairs.  We write
\[
\ip_i^{\refl}
:=
\operatorname{inc}_i\,\operatorname{pre}_{\bar i},
\]
so that
\[
\ip_{\bar i}^{\refl}
:=
\operatorname{inc}_{\bar i}\,\operatorname{pre}_i.
\]
Thus, when the reflected pair $\{i,\bar i\}$ is added to the boundary, one chooses on which side the coloop is placed; the reflected side is then forced to be a loop.  These pair insertions preserve the Lagrangian symmetry and will be used below to build the Lagrangian Grassmannian cells.

The second operation is the positive paired bridge.  Write a matrix representative as
\[
C=(c_1,\ldots,c_{2n}),
\]
where $c_i$ is the $i$th column.  Away from the reflection axes, the right bridge at $i$ acts by
\[
c_{i+1}\longmapsto c_{i+1}+w c_i,
\qquad
c_{\bar i}\longmapsto c_{\bar i}+w c_{\overline{i+1}},
\qquad
w>0.
\]
Equivalently, this is right multiplication by
\begin{equation}\label{eq:right_refl_bridge}
M_i^{\refl,\rightarrow}(w)
=
I+wX_i,
\qquad
X_i=\mathsf e_{i,i+1}+\mathsf e_{\overline{i+1},\bar i}.
\end{equation}
The reflected left bridge is obtained by transposition,
\begin{equation}\label{eq:left_refl_bridge}
M_i^{\refl,\leftarrow}(w)
=
\bigl(M_i^{\refl,\rightarrow}(w)\bigr)^{\mathsf T}.
\end{equation}
At the middle reflection axis there is only one bridge, whose right version is
\begin{equation}\label{eq:refl_bridge_mid}
M_{\mathrm{mid}}^{\refl,\rightarrow}(w)
=
I+w\mathsf e_{n,n+1},
\end{equation}and the left version is similar.
At the outer cyclic axis, if we keep the fixed linear order
$1<2<\cdots<2n$ for Pl\"ucker coordinates, the positivity-preserving normalization is
\begin{equation}\label{eq:refl_bridge_out}
M_{\mathrm{out}}^{\refl,\rightarrow}(w)
=
I+(-1)^{n-1}w\mathsf e_{2n,1}.
\end{equation}
The sign is the usual cyclic sign from the ordinary positive Grassmannian convention.

All these matrices are symplectic for Karpman's form.  Namely,
\[
M E_{\refl} M^{\mathsf T}=E_{\refl}.
\]
Therefore, if $C\in\LG^{\refl}(n,2n)$, then also
\[
CM\in\LG^{\refl}(n,2n).
\]
For $w>0$, these operations are subtraction-free in the corresponding positive charts.  In particular, they preserve nonnegativity, and positive bridge words increase positivity by turning on controlled families of Pl\"ucker coordinates which were zero before the bridge was attached.

\begin{remark}[Evolving label sets]
\label{rem:evolving-label-sets}
The operations \(\inc_i\), \(\pre_i\), and hence
\(\ip_i^{\refl}\), should be understood as operations on cells with
\emph{evolving label sets}.  Thus, if \(S\subset[2n]\) is a
reflection-stable set of labels and the pair \(\{i,\bar i\}\) is not yet in
\(S\), then \(\ip_i^{\refl}\) maps a reflected Lagrangian cell on \(S\) to a
reflected Lagrangian cell on
\[
S\sqcup\{i,\bar i\},
\]
after inserting the two new labels in their cyclic positions.  
%
One way to construct a reflected Lagrangian cell is by performing a series of \(\ip\)-operations and paired bridge operations.
\end{remark}

\section{Symplectic kinematics and the  Lagrangian amplituhedron}
\label{sec:symplectic-kinematics}
We now pass from the domain Grassmannians to the main construction of the paper: the Lagrangian amplituhedron and the induced spinor-helicity kinematics. 

Write
\[
\Omega_\refl:=E_\refl,
\]
then
\[{\LG^\refl_{\geq}(n,2n)}
=
\{\,C\in\Gr_{\geq}(n,2n):
{C\Omega_\refl C^{\mathsf T}=0\,\}.}
\]
\subsection{The amplituhedron maps}

Let
\[
W\in\Mat(2n,n+2)
\]
be a full-rank external matrix, with rows \(W_1,\ldots,W_{2n}\).  We call \(W\) \emph{positive} if its ordered maximal minors are positive, equivalently if
\[
\operatorname{colspan}(W)\in\Gr_{>}(n+2,2n).
\]
Later, when discussing signs of Mandelstam variables, we will sometimes impose stronger positivity assumptions on \(W\).

For
\[
{C\in\LG^\refl_{\geq}(n,2n),}
\]
define
\[
Y=CW\in\Gr(n,n+2).
\]
{The corresponding (reflected) Lagrangian amplituhedron is}
\[
{\Lampl_n^\refl(W)}
=
\{\,\operatorname{rowspan}(CW):
{C\in\LG^\refl_{\geq}(n,2n)\,\}}
\subset\Gr(n,n+2).
\]
Here \(W\) denotes the external data of the Lagrangian amplituhedron.  We use \(W\), rather than \(\Lambda\) or \(Z\), to avoid confusion with the physical spinors \(\lambda,\widetilde\lambda\) and with momentum-twistor notation.

Let
\[
\mathcal W:=\operatorname{colspan}(W)\subset\mathbb R^{2n}.
\]
The natural \(B\)-picture two-plane is
\[
\widetilde\lambda
=
C^\perp\cap\mathcal W.
\]
This notation is chosen to match the standard momentum-amplituhedron convention recalled in the background: the spinors lying in \(C^\perp\) are denoted by \(\widetilde\lambda\), while the chiral spinors \(\lambda\) will lie in \(C\).  This convention is used only for the Lagrangian amplituhedra and does not change the notation of the ordinary momentum amplituhedron.

For generic \(C\) and \(W\),
\[
\dim(C^\perp\cap\mathcal W)
=
 n+(n+2)-2n
=
2.
\]Karp shows \cite{karp2017sign} that under the positivity assumptions we use the intersection admits the expected dimension.
We choose a \(2\times 2n\) matrix representative, again denoted by
\[
\widetilde\lambda
=(\widetilde\lambda_1,\ldots,\widetilde\lambda_{2n}),
\qquad
\widetilde\lambda_i\in\mathbb R^2.
\]
The two-by-two square brackets are
\[
[ij]:=\det(\widetilde\lambda_i,\widetilde\lambda_j).
\]

This description is equivalent to the usual projected \(Y\)-space description.  Indeed, if \(K\) is an \((n+2)\times2\) matrix whose columns span the kernel of \(Y=CW\), then
\[
WK\subset C^\perp\cap\mathcal W,
\]
and generically \(\widetilde\lambda=(WK)^{\mathsf T}\), after a choice of basis of the two-plane.  Consequently,
\[
[ij]
=
\det
\begin{pmatrix}
Y\\
W_i\\
W_j
\end{pmatrix},
\]
up to the common normalization determined by the chosen bases.

\subsection{The symplectic constraint and momentum conservation}

{The Lagrangian condition implies}
\[
{C^\perp=\operatorname{rowspan}(C\Omega_\refl)}
\]
with respect to the ordinary Euclidean pairing.  Therefore the two-plane
\(\widetilde\lambda\subset C^\perp\) satisfies
\[
{\widetilde\lambda\Omega_\refl\widetilde\lambda^{\mathsf T}=0.}
\]
This is the basic equation behind the folded kinematics.

{We define the chiral spinors from the \(B\)-plane by applying the symplectic form}
\[
\lambda=
\widetilde\lambda \Omega_\refl=\widetilde\lambda E_\refl.
\]
{Since \(C^\perp=C\Omega_\refl\) and \(\Omega_\refl^2=-I\), the chiral plane \(\lambda\) lies in \(C\). Thus}
\[
\lambda\subset C,
\qquad
\widetilde\lambda\subset C^\perp,
\]
exactly as in the standard momentum-amplituhedron convention.

{We define rank-one momentum matrices by}
\[
p_i=\lambda_i\widetilde\lambda_i^{\mathsf T}.
\]
Then
\[
\sum_{i=1}^{2n}p_i
=
\lambda\widetilde\lambda^{\mathsf T}
\]
{is a scalar multiple of \(\widetilde\lambda\Omega_\refl\widetilde\lambda^{\mathsf T}\).  Hence}
\[
\sum_{i=1}^{2n}p_i=0.
\]
Thus the symplectic isotropy of the \(B\)-plane is precisely momentum conservation.

The ordinary little-group action scales
\[
\lambda_i\longmapsto t_i\lambda_i,
\qquad
\widetilde\lambda_i\longmapsto t_i^{-1}\widetilde\lambda_i.
\]
The folded relations below force the parameters on paired labels to be inverse to one another.  Thus the compatible little group is isomorphic to \((\mathbb C^\times)^n\).

\subsection{Mandelstam variables and folded symmetries}

{For any subset \(I\subset[2n]\), define}
\[
P_I:=\sum_{i\in I}p_i,
\qquad
s_I:=\det(P_I).
\]
Equivalently,
\[
s_I
=
\sum_{\substack{i<j\\ i,j\in I}}
\langle ij\rangle[ij],
\qquad
\langle ij\rangle:=\det(\lambda_i,\lambda_j).
\]

We have\[
p_{\bar{i}}=-p_i^{\mathsf T}.
\]
Hence
\[
P_{\bar{I}}=-P_I^{\mathsf T}.
\]
Momentum conservation gives
\[
P_{I^c}=-P_I.
\]
Therefore the Mandelstam variables satisfy the Klein-four symmetry
\begin{equation}
\label{eq:klein-mandelstam}
{s_I=s_{\bar{I}}=s_{I^c}=s_{\overline{I^c}}.}
\end{equation}
Thus, when studying interval Mandelstams, we may quotient by the involution \(\tau\) and by complementation.

The importance of Mandelstam variables when studying spinor-helicity amplituhedra is that their zero loci, or subsets thereof, are conjecturally their (real) codimension $1$ boundaries. This is proven, under stronger positivity conditions, for the momentum and ABJM amplituhedra \cite{Galashin2024AmplituhedraOrigami,OrenPerlsteinTessler2025ABJMBCFWTiling}.
\subsection{The reflected specialization}

In the reflected model the involution is
\[
\bar i:=2n+1-i.
\]
Since
\[
(E_\refl)_{ji}=(-1)^i\delta_{j,\bar i},
\]
the definition
\[
\lambda=\widetilde\lambda E_\refl
\]
is equivalently
\begin{equation}\label{eq:lambd_tilde_lambd}
\lambda_i=(-1)^i\widetilde\lambda_{\bar i}.
\end{equation}
Therefore
\begin{equation}\label{eq:p_i_bar_i}
p_i=(-1)^i\widetilde\lambda_{\bar i}\widetilde\lambda_i^{\mathsf T},
\end{equation}
and, because \((-1)^{\bar i}=-(-1)^i\), we get
\[
p_{\bar i}=-p_i^{\mathsf T}.
\]
The compatible little group is
\[
t_{\bar i}=t_i^{-1}.
\]
The angle brackets are folded square brackets:
\[
\langle ij\rangle
=
(-1)^{i+j}[\bar i\,\bar j].
\]
The scalar form of momentum conservation is
\[
\sum_{i=1}^n(-1)^{i+1}[i\,\bar i]=0.
\]
For later use we define
\begin{equation}\label{eq:mu}
\mu_r
:=
\sum_{i=1}^r(-1)^{i+1}[i\,\bar i],
\qquad
0\leq r\leq n,
\end{equation}
so that
\[
\mu_0=\mu_n=0.
\]
Let
\begin{equation}\label{eq:J_r}
J_r:=[\bar r,r]_{\cyc}
=
\{\bar r,\ldots,\bar1,1,\ldots,r\}.
\end{equation}
Then
\[
P_{J_r}
=
\sum_{i\in J_r}p_i
=
\mu_r
\begin{pmatrix}
0&1\\
-1&0
\end{pmatrix},
\]
and hence
\begin{equation}\label{eq:mu_square}
s_{J_r}=\det(P_{J_r})=\mu_r^2.
\end{equation}
In particular, the vanishing of \(s_{J_r}\) is equivalent to
\[
\mu_r=0.
\]
For an interval contained in one half,
\[
I=[a,b]\subseteq[1,n],
\]
the Mandelstam variable takes the form
\[
s_I
=
\sum_{a\leq i<j\leq b}
(-1)^{i+j+1}
[ij]
[\bar j\,\bar i],
\]
where the second bracket is written in increasing order.  In particular, for a two-particle interval contained in one half,
\[
s_{[a,a+1]}
=
[a,a+1]
[\overline{a+1},\bar a].
\]
Thus the two-particle Mandelstam equation has the two collinear components
\[
[a,a+1]=0
\qquad\text{or}\qquad
[\overline{a+1},\bar a]=0.
\]
The two length-two intervals which cross the reflection axes are already covered by the symmetric variables: the outer one is \(J_1=[\bar1,1]_{\cyc}\), while the middle one is the complement of \(J_{n-1}\).  Their Mandelstams are therefore \(\mu_1^2\) and \(\mu_{n-1}^2\), respectively.

It remains to discuss cyclic intervals which are neither symmetric nor contained in one half.  Set
\[
Q_a:=\sum_{i=1}^a p_i,
\qquad
1\leq a\leq n,
\]
and consider the axis-crossing interval
\[
J_{a,b}:=[\bar b,a]_{\cyc}.
\]
where $1\leq b\leq n.$ Then
\[
P_{J_{a,b}}=Q_a-Q_b^{\mathsf T}.
\]
A direct \(2\times2\) determinant computation gives
\begin{equation}\label{eq:asymmetric-Mandelstam-identity-strong-section-general}
s_{J_{a,b}}
=
s_{[\min(a,b)+1,\max(a,b)]}
+
\mu_a\mu_b,
\end{equation}
with the conventions
\[
s_{\varnothing}=s_{\{i\}}=0.
\]
In particular, for \(a<b\),
\begin{equation}\label{eq:asymmetric-Mandelstam-identity-strong-section}
s_{[\bar b,a]_{\cyc}}
=
s_{[a+1,b]}+\mu_a\mu_b.
\end{equation}
There is one adjacent degeneration.  If \(b=a+1\), then
\begin{equation}\label{eq:asymmetric-Mandelstam-identity-strong-section-special}
s_{[\overline{a+1},a]_{\cyc}}
=
\mu_a\mu_{a+1}.
\end{equation}
Thus its zero locus is the union of two already existing symmetric supports.

For \(b\geq a+2\), and under the strong sign assumptions we shall define later, the two summands
\[s_{[a+1,b]},\qquad\mu_a\mu_b\]
have the same sign.  Hence the vanishing of
\[s_{[\bar b,a]_{\cyc}}=s_{[a+1,b]}+\mu_a\mu_b,\]at least under the stronger conditions,
forces the simultaneous vanishing of one half-contained Mandelstam and one symmetric variable.  Thus these asymmetric interval Mandelstams do not give codimension-one boundary supports; they are expected to vanish only on higher-codimension intersections of the primary supports.

Consequently, modulo reflection and complement, the functionaries whose zero loci are expected to contain the real codimension-one boundary components of the reflected Lagrangian amplituhedron are
\[
\mu_r=0,
\qquad
1\leq r\leq n-1,
\]
and
\[
s_{[a,b]}=0,
\qquad
1\leq a<b\leq n.
\]

\subsection{Bridge promotion and boundary functionaries}
\label{subsec:bridge-promotion-principle}
We record a general mechanism for studying the effect of positive bridge operations on the sign or vanishing of functionaries.  This is the Lagrangian analog of the \emph{promotion principle} developed in the proof of BCFW tiling conjecture for $\Ampl_{n,k,4}$ \cite[Section 4]{EvenZoharLakrecTessler2025BCFWTriangulation}, and we refer the reader there for precise details.  
Let
\[
{M=M^\refl(t),}
\qquad t>0,
\]
be one of the positivity-preserving bridge matrices of \cref{subsub:refl_positive_ops}.
\[
M\Omega_\refl M^{\mathsf T}=\Omega_\refl,
\]
and right multiplication gives a positive domain operation
\[
C\longmapsto CM.
\]
The same matrix can also be applied to the external datum,
\[
W\longmapsto MW.
\]
The two viewpoints are related by the transfer identity
\begin{equation}
\label{eq:bridge-transfer-identity}
C(MW)=(CM)W.
\end{equation}
Thus the same point of the image may be represented either by the bridged
domain point \(CM\) with external data \(W\), or by the unbridged point \(C\)
with shifted external data \(MW\).

{Let \(\mathcal S\subset\LG^\refl_{\geq}(n,2n)\) be a nonnegative domain cell, and let}
\[
\mathcal S'=\mathcal S\cdot M(\mathbb R_{>0})
\]
be the cell obtained by adding the bridge.  Let \(\mathfrak F\) be a functionary, for
example a Mandelstam variable, a spinor bracket, or a polynomial expression in them.  If
\[
C'=CM(t)\in\mathcal S',
\]
then, as above,
\[
C'W
=
C\bigl(M(t)W\bigr).
\]

The bridge therefore defines a promoted functionary on the \emph{bridged cell} by
\begin{equation}
\label{eq:bridge-promoted-function}
\mathfrak F^M_t(C';W)
:=
\mathfrak F\bigl(C'M(-t);M(t)W\bigr).
\end{equation}
Equivalently, if \(C'=CM(t)\), then
\[
\mathfrak F^M_t(CM(t);W)
=
\mathfrak F(C;M(t)W).
\]
This is the promotion principle we will use.  

Consequently, if \(\mathfrak F(C;W)\) vanishes, or has a fixed sign, for all
\(C\in\mathcal S\) and all positive external data \(W\), then the promoted
functionary \(\mathfrak F^M_t\) has the same vanishing or sign property on
\(\mathcal S'\): Indeed, for \(t>0\), the shifted datum \(M(t)W\) is again
positive.

The same construction is also a practical way to undo a bridge.  Suppose that
\[
C'\in\mathcal S'
\]
and let \(t\) be a candidate bridge-removal parameter.  Set
\[
C(t):=C'M(-t).
\]
Then
\begin{equation}
\label{eq:bridge-removal-transfer}
C'W=C(t)\bigl(M(t)W\bigr).
\end{equation}
Therefore a boundary-defining equation on the unbridged cell, for example the
vanishing of a Mandelstam variable, becomes an equation in \(t\) after moving
the bridge from the domain to the external data.

Let
\(
\widetilde\lambda,\lambda
\)
be the spinor variables corresponding to the bridged presentation \((C',W)\),
so that
\[
\widetilde\lambda\subset (C')^\perp,
\qquad
\lambda\subset C'.
\]
When the bridge is moved to the external data as in
\eqref{eq:bridge-removal-transfer}, the spinor variables of the unbridged
presentation \((C(t),M(t)W)\) are obtained by the substitution
\begin{equation}
\label{eq:bridge-spinor-substitution}
\widetilde\lambda(t)=\widetilde\lambda M(t)^{\mathsf T},
\qquad
\lambda(t)=\lambda M(t)^{-1}.
\end{equation}
Equivalently, in column-vector notation, \(\widetilde\lambda\) is acted on by
\(M(t)\), while \(\lambda\) is acted on by \(M(t)^{-\mathsf T}\).

Let us justify \eqref{eq:bridge-spinor-substitution}.  If \(K\) spans the
kernel of the fixed target plane
\[
Y=C'W,
\]
then
\[
\widetilde\lambda=(WK)^{\mathsf T}.
\]
Replacing \(W\) by \(M(t)W\) gives
\[
(M(t)WK)^{\mathsf T}
=
(WK)^{\mathsf T}M(t)^{\mathsf T},
\]
which proves the formula for \(\widetilde\lambda(t)\). The proof for $\lambda(t)$ is similar. 
Note that,\[\lambda(t)\widetilde\lambda(t)^{\mathsf T}=\lambda M(t)^{-1}M(t)\widetilde\lambda^{\mathsf T}=\lambda\widetilde\lambda^{\mathsf T},\]
so momentum conservation is preserved.

Thus, if a functionary has a spinor expression
\[
F(\widetilde\lambda,\lambda),
\]
then its bridge-promoted version on the bridged cell is obtained by expanding
\begin{equation}
\label{eq:bridge-expanded-function}
F_t
:=
F\bigl(
\widetilde\lambda M(t)^{\mathsf T},
\lambda M(t)^{-1}
\bigr).
\end{equation}
There are two cases which will appear repeatedly.

First, \(F_t\) may be independent of \(t\).  Then the promoted functionary has
the same sign or vanishing property throughout the bridged cell.  This is the
mechanism used below when constructing domain cells contained in zero loci
such as
\[
P_I=0,
\qquad
\mu_r=0,
\qquad
s_I=0.
\]

Second, the parameter \(t\) may appear nontrivially.  On a cell whose
amplituhedron map is finite-to-one, \(t\) can then be recovered locally from
the target spinors by solving
\begin{equation}
\label{eq:bridge-root-equation}
F_t=0.
\end{equation}
{This is the mechanism used in the BCFW recursion.  In the reflected Lagrangian model studied here the equations needed below are linear in \(t\).  This is in contrast to the orthogonal (ABJM) model, where equations are usually quadratic.}

\section{Boundaries, BCFW recursion, and a semi algebraic description}
\label{sec:reflected-amplituhedron}

In this section we study the boundary structure of the reflected Lagrangian amplituhedron.  The domain is Karpman's nonnegative Lagrangian Grassmannian
\[
\LG_{\geq}^{\refl}(n,2n)\subset \Gr_{\geq}(n,2n),
\]
with reflection involution
\[
\bar i=2n+1-i.
\]
The amplituhedron map is
\[
\Phi_W:\LG_{\geq}^{\refl}(n,2n)\longrightarrow \Gr(n,n+2),
\qquad
C\longmapsto CW,
\]
where
\[
W\in\Mat(2n,n+2)
\]
is the reflected Lagrangian external datum.

Recall from \cref{sec:symplectic-kinematics} that the primary reflected boundary functionaries are of two types.  The first are the reduced symmetric functions
\[
\mu_r,
\qquad
1\leq r\leq n-1,
\]
associated with the reflection-invariant intervals
\[
J_r=[\bar r,r]_{\cyc},
\qquad
s_{J_r}=\mu_r^2.
\]
The second are the half-contained Mandelstams
\[
s_I,
\qquad
I=[a,b]\subseteq[1,n].
\]

The maximal \emph{boundary cells}, meaning cells of \(\LG_{\geq}^{\refl}(n,2n)\) whose images lie in the zero loci of these boundary functionaries, will be described below.  When the external data satisfy the strong positivity assumptions, one can show that these images are true codimension-one boundary strata, and that their closures are exactly the intersection between the ampliuthedron and the zero loci.  Here we explain why the images lie in the respective zero loci.  Maximality is a consequence of a more detailed study of Temperley--Lieb immanant expansions, and will be verified in \cite{MS}.
\subsection{Symmetric boundaries: \(\mu_r=0\)}
\label{subsec:reflected-symmetric-boundaries}

Fix
\[
1\leq r\leq n-1
\]
and define
\[
J_r=[\bar r,r]_{\cyc}
=
\{\bar r,\ldots,\bar1,1,\ldots,r\}.
\]
Its complement is
\[
J_r^c=[r+1,\overline{r+1}]_{\cyc}.
\]
Both \(J_r\) and \(J_r^c\) are stable under the reflection involution.  The corresponding Mandelstam variable is a square,
\[
s_{J_r}=\mu_r^2,
\]
so the reduced boundary equation is
\[
\mu_r=0.
\]

\subsubsection{The maximal block cell}

The maximal domain cell over \(\mu_r=0\) is the block Lagrangian locus
\[
C=C_1\oplus C_2,
\]
where
\[
C_1\in\LG_{\geq}^{\refl}(r,2r)
\]
is supported on \(J_r\), and
\[
C_2\in\LG_{\geq}^{\refl}(n-r,2n-2r)
\]
is supported on \(J_r^c\).  After ordering the columns according to the decomposition
\[
[2n]=J_r\sqcup J_r^c,
\]
a representative has the form
\[
C=
\begin{pmatrix}
C_1&0\\
0&C_2
\end{pmatrix}.
\]
Since both blocks are Lagrangian for the restricted Karpman forms, the direct sum is Lagrangian for \(E_{\refl}\).  Since both blocks are nonnegative, the direct sum lies in
\[
\LG_{\geq}^{\refl}(n,2n).
\]
The domain dimension of this cell is
\[
d_{\mathrm{sym}}(r)
=
\frac{r(r+1)}2+
\frac{(n-r)(n-r+1)}2.
\]
Equivalently, its codimension inside the full reflected Lagrangian Grassmannian is
\[
\frac{n(n+1)}2-d_{\mathrm{sym}}(r)
=
r(n-r).
\]

\subsubsection{Why \(\mu_r\) vanishes}

For a point of the block cell, the restricted spinor data on \(J_r\) form a reflected Lagrangian kinematical system.  Hence
\[
P_{J_r}=0.
\]
But in the reflected kinematics
\[
P_{J_r}
=
\mu_r
\begin{pmatrix}
0&1\\
-1&0
\end{pmatrix}.
\]
Therefore
\[
\mu_r=0.
\]
Thus the block cell is contained in the reduced symmetric boundary.  By total momentum conservation, the complementary subsystem \(J_r^c\) is separately momentum-conserving as well.

\subsubsection{Construction by positive operations}

The same cell can be constructed using Karpman's elementary positive operations.  Start with the top cell
\[
\LG_{>}^{\refl}(r,J_r)
\]
supported on the reflected label set \(J_r\).  The missing reflected pairs are
\[
\{i,\bar i\},
\qquad
i=r+1,\ldots,n.
\]
Insert them by reflected loop-coloop operations.  In one convenient cyclic chart this can be written as the word
\[
\ip_n^{\refl},\ \ip_{n-1}^{\refl},\ \ldots,\ \ip_{r+1}^{\refl},
\]
where $\ip$ are the operations from \cref{subsub:refl_positive_ops}.
Then apply arbitrary bridges \eqref{eq:right_refl_bridge},~\eqref{eq:left_refl_bridge},~\eqref{eq:refl_bridge_out} which not affect the the elements in $J_r^c,$ until the big Lagrangian cell of $\LG_\geq^\refl(r,J_r)$ is attained. These operations clearly do not affect $\lambda_i,\tilde\lambda_i$ for $i\in J_r^c,$ hence also do not affect $s_{J_r^c},$ which is initially $0.$ Thus, in the end of the process $s_{J_r^c}=0,$ hence also $s_{J_r}=0,$ which implies $\mu_r=0$, by \eqref{eq:mu_square}, on the image of the resulting domain cell.


\subsubsection{Karpman graph picture}

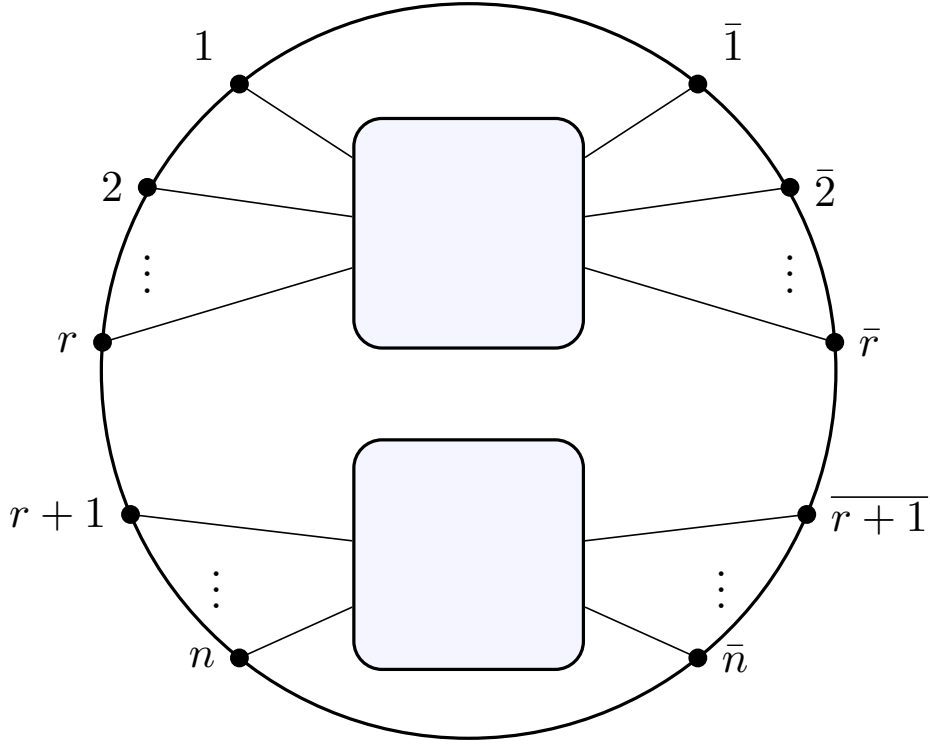
\begin{figure}[h!]
\centering
\resizebox{0.8\textwidth}{!}{%
\begin{tikzpicture}
\begin{feynman}
  \draw[thick] (0,0) circle (3.2);

  \vertex[blob, minimum size=2cm] (T) at (0,1.2) {};
  \vertex[blob, minimum size=2cm] (B) at (0,-1.6) {};

  \vertex[dot, label={[label distance=1pt]above left:$1$}] (L1) at (-1.997, 2.50) {};
  \vertex[dot, label=left:$2$] (L2) at (-2.8, 1.6) {};
  \vertex[dot, label=left:$r$] (L3) at (-3.190, 0.25) {};
  \vertex[dot, label=left:$r+1$] (L4) at (-2.946, -1.25) {};
  \vertex[dot, label=left:$n$] (L5) at (-1.997, -2.50) {};

  \vertex[dot, label={[label distance=1pt]above right:$\bar{1}$}] (R1) at (1.997, 2.50) {};
  \vertex[dot, label=right:$\bar{2}$] (R2) at (2.8, 1.6) {};
  \vertex[dot, label=right:$\bar{r}$] (R3) at (3.190, 0.25) {};
  \vertex[dot, label=right:$\overline{r+1}$] (R4) at (2.946, -1.25) {};
  \vertex[dot, label=right:$\bar{n}$] (R5) at (1.997, -2.50) {};

  \diagram*{
    (L1) -- (T), (L2) -- (T), (L3) -- (T),
    (L4) -- (B), (L5) -- (B),
    (R1) -- (T), (R2) -- (T), (R3) -- (T),
    (R4) -- (B), (R5) -- (B),
  };

  \node at (-2.80, 0.95) {$\vdots$};
  \node at (-2.2, -1.8) {$\vdots$};
  \node at (2.80, 0.95) {$\vdots$};
  \node at (2.2, -1.8) {$\vdots$};
\end{feynman}
\end{tikzpicture}
}
\caption{The plabic tangle for symmetric boundaries.}
\label{fig:diagram1}
\end{figure}

At the level of Karpman graphs, the symmetric boundary is obtained by separating the graph into two reflected components.  One component has boundary labels \(J_r\) and represents a cell of
\[
\LG_{\geq}^{\refl}(r,2r),
\]
while the other has boundary labels \(J_r^c\) and represents a cell of
\[
\LG_{\geq}^{\refl}(n-r,2n-2r).
\]
Each component is itself a symmetric plabic graph: reflection preserves the underlying uncolored graph and swaps colors.  No bridge crosses the cut between the two components.
See \cref{fig:diagram1}.

Thus the domain and graph structure of the symmetric boundary is
\[
\LG_{\geq}^{\refl}(r,2r)
\times
\LG_{\geq}^{\refl}(n-r,2n-2r).
\]
This product description is literal at the domain Grassmannian and graph level; we do not mean that the image boundary in \(\Gr(n,n+2)\) is a na\"ive Cartesian product.
\subsection{Half-interval boundaries: \(s_{[a,b]}=0\)}
\label{subsec:reflected-half-interval-boundaries}

Let
\[
I=[a,b]\subseteq[1,n],
\qquad
m:=|I|=b-a+1,
\]
and write
\[
\bar I=[\bar b,\bar a]
\]
for the reflected interval, and define the reflection-stable complement
\[
K=[2n]\setminus(I\sqcup\bar I).
\]
The boundary over
\[
s_I=0
\]
has a different structure from the symmetric boundary.  It is not a product of two smaller reflected Lagrangian systems.  Instead, its maximal cells are built from an ordinary momentum-amplituhedron blob, a smaller reflected Lagrangian core, and the reflected color-dual ordinary blob.

\subsubsection{The maximal cells}

The maximal cells over \(s_I=0\) are indexed by
\[
r,s\geq0,
\qquad
r+s=m-1.
\]
A construction recipe for the maximal cells in the zero locus of $s_I$ is given by the following sequence of operations. 
Start with the top reflected Lagrangian cell on the reflection-stable label set
\[
\bigl([n]\setminus[a,b-1]\bigr)
\sqcup
\overline{\bigl([n]\setminus[a,b-1]\bigr)}.
\]
There are \(m-1\) missing reflected pairs,
\[
\{i,\bar i\},
\qquad
i=a,\ldots,b-1.
\]
Insert them using \(m-1\) operations of type \(\ip^{\refl}\).  For each missing pair, choose whether the coloop lies on the \(I\)-side or on the \(\bar I\)-side.  Suppose the coloop lies on the \(I\)-side exactly \(r\) times and on the \(\bar I\)-side exactly \(s\) times.  After applying enough positivity-improving bridges \eqref{eq:right_refl_bridge},~\eqref{eq:left_refl_bridge} internal to the blobs, one obtains the maximal cell associated with the pair
\[
(r,s).
\]
Note that among the newly added $r+s$ rows, exactly $r$ rows have the property that their restriction to the entries of $I$ are non zero, and exactly $s$ have the analogous property with respect to $\bar{I}.$

This construction uses only Karpman's positivity-preserving operations: reflected loop-coloop insertions and reflected bridge operations.  It also makes the reflection constraint transparent: the two ordinary blobs are not independent; one is the reflected color-dual of the other.

The bridge-promotion principle of \cref{subsec:bridge-promotion-principle} explains why the subsequent internal bridges do not move the cell out of the Mandelstam zero locus.  Once the ordinary blob, the reflected color-dual blob, and the smaller reflected core have been assembled, the allowed positivity-improving bridges are internal to these pieces.  In the bridge substitution \eqref{eq:bridge-spinor-substitution}, such internal bridges do not change the partial momentum \(P_I\) across the cut defining the half-interval channel.  Thus the equation
\[
s_I=\det(P_I)=0
\]
is preserved under the remaining positive operations.

\subsubsection{Linear-algebra preparation}

Choose rows so that a point of the \((r,s)\)-cell is described by three pieces.  First, there is an \(r\)-dimensional row subspace
\[
L\subset\mathbb R^I
\]
supported on \(I\).  Second, there is an \(s\)-dimensional row subspace
\[
R\subset\mathbb R^{\bar I}
\]
supported on \(\bar I\).  Third, there is a complementary row subspace
\[
H
\]
of dimension
\[
h:=n-r-s=n-m+1.
\]
Rows in \(H\) may have components on \(I\), on \(K\), and on \(\bar I\).

To compare the two outer blobs, define the signed reversal
\[
\sigma:\mathbb R^{\bar I}\longrightarrow \mathbb R^I,
\qquad
(\sigma y)_i=(-1)^{i+1}y_{\bar i}.
\]
This convention is chosen so that for
\(
x\in\mathbb R^I,
\qquad
y\in\mathbb R^{\bar I},
\)
one has
\[
xE_{\refl}y^{\mathsf T}=x\cdot\sigma(y).
\]
Set
\[
\widetilde L:=\sigma(R)\subset\mathbb R^I.
\]
The symplectic condition implies
\[
L\perp\widetilde L
\]
with respect to the ordinary dot product on \(\mathbb R^I\).

Let \(\lambda_I\) and \(\widetilde\lambda_I\) be the restrictions of the two spinor planes to the interval \(I\).  The incidence relations of the cell are
\[
L\perp\widetilde\lambda_I,
\qquad
\widetilde L\perp\lambda_I,
\qquad
L\perp\widetilde L.
\]
Since
\[
m=r+s+1,
\]
generic dimension counting gives
\[
\dim(\lambda_I\cap L)=1,
\qquad
\dim(\widetilde\lambda_I\cap\widetilde L)=1.
\]
Choose nonzero vectors
\[
v_\lambda\in\lambda_I\cap L,
\qquad
v_{\widetilde\lambda}\in\widetilde\lambda_I\cap\widetilde L.
\]

\subsubsection{The internal particle and the ordinary momentum blob}
Since $s_I=0$ the partial momentum matrix
\[
P_I=\lambda_I\widetilde\lambda_I^{\mathsf T}
\] generically has rank $1.$
At a generic point of \(s_I=0\), choose internal spinors
\[
\lambda_{\st},\widetilde\lambda_{\st}\in\mathbb R^2
\]
such that
\[
P_I=-\lambda_{\st}\widetilde\lambda_{\st}^{\mathsf T}.
\]
They are unique up to the usual little-group scaling
\[
(\lambda_{\st},\widetilde\lambda_{\st})
\longmapsto
(t\lambda_{\st},t^{-1}\widetilde\lambda_{\st}).
\]
Define extended spinor configurations on \(I\cup\{\st\}\) by
\[
\lambda^{\mathrm{ext}}
=
(\lambda_I\mid\lambda_{\st}),
\qquad
\widetilde\lambda^{\mathrm{ext}}
=
(\widetilde\lambda_I\mid\widetilde\lambda_{\st}).
\]
Then
\[
\lambda^{\mathrm{ext}}
(\widetilde\lambda^{\mathrm{ext}})^{\mathsf T}=0.
\]
The vectors \(v_\lambda\) and \(v_{\widetilde\lambda}\) extend with zero internal coordinate:
\[
(v_\lambda,0)\in\lambda^{\mathrm{ext}},
\qquad
(v_{\widetilde\lambda},0)\in\widetilde\lambda^{\mathrm{ext}}.
\]
Choose complementary basis vectors and normalize their internal entries to one:
\[
u_\lambda=(q_{\st},1),
\qquad
u_{\widetilde\lambda}=(q_{\bar\st},1).
\]
Thus
\[
\lambda^{\mathrm{ext}}
=
\operatorname{span}\{(v_\lambda,0),u_\lambda\},
\]
and
\[
\widetilde\lambda^{\mathrm{ext}}
=
\operatorname{span}\{(v_{\widetilde\lambda},0),u_{\widetilde\lambda}\}.
\]
Insert a zero \(\st\)-column into \(L\) and \(\widetilde L\), and define
\[
L^{\mathrm{ext}}
=
(L\oplus0)+\langle u_\lambda\rangle,
\]
\[
\widetilde L^{\mathrm{ext}}
=
(\widetilde L\oplus0)+\langle u_{\widetilde\lambda}\rangle.
\]
Then
\[
\dim L^{\mathrm{ext}}=r+1,
\qquad
\dim\widetilde L^{\mathrm{ext}}=s+1,
\]
and
\[
L^{\mathrm{ext}}\perp\widetilde L^{\mathrm{ext}}.
\]
Since
\[
(r+1)+(s+1)=m+1,
\]
we have
\[
\widetilde L^{\mathrm{ext}}=(L^{\mathrm{ext}})^\perp.
\]

With the standard momentum-amplituhedron convention used in this paper, the ordinary domain plane is
\[
C_{\mathrm{mom}}
:=
L^{\mathrm{ext}}
\in\Gr_{\geq}(r+1,m+1).
\]
Indeed, the incidence relations become
\[
\lambda^{\mathrm{ext}}\subset C_{\mathrm{mom}},
\qquad
\widetilde\lambda^{\mathrm{ext}}\subset C_{\mathrm{mom}}^\perp.
\]
Thus the ordinary blob is a momentum-amplituhedron factor of type
\[
\mathcal M_{m+1,r+1}.
\]
The reflected blob is then forced to be the reflected color-dual ordinary factor, with complementary sector \(s+1\).

\subsubsection{The  reflected Lagrangian core}

We now describe the remaining reflected component.  For a vector \(h\in H\), isotropy with the two outer blobs implies
\[
h_I\in\widetilde L^\perp,
\qquad
\sigma(h_{\bar I})\in L^\perp,
\]
where $h_A$ is the restriction of $h$ to the label set $A.$ Generically,
\[
\widetilde L^\perp=L\oplus\langle q_{\st}\rangle,
\qquad
L^\perp=\widetilde L\oplus\langle q_{\bar\st}\rangle.
\]
Thus there are unique coefficients \(x_{\st}(h)\) and \(x_{\bar\st}(h)\), modulo rows of \(L\oplus R\), such that
\[
h_I=\ell(h)+x_{\st}(h)q_{\st},
\qquad
\ell(h)\in L,
\]
and
\[
\sigma(h_{\bar I})
=
\widetilde\ell(h)+x_{\bar\st}(h)q_{\bar\st},
\qquad
\widetilde\ell(h)\in\widetilde L.
\]
Define
\[
H^{\mathrm{ext}}
=
\left\{
\bigl(x_{\st}(h),h_K,x_{\bar\st}(h)\bigr):h\in H
\right\}
\subset
\mathbb R^{\{\st\}\sqcup K\sqcup\{\bar\st\}}.
\]
This space has dimension
\[
h=n-m+1.
\]
The ambient label set
\[
\{\st\}\sqcup K\sqcup\{\bar\st\}
\]
has size \(2h\), and it carries the induced reflected Karpman form, with \(\st\) and \(\bar\st\) forming a new reflected pair.  A direct substitution into the symplectic pairing shows that
\[
H^{\mathrm{ext}}\in\LG^{\refl}(h,2h).
\]
The rescaling
\[
q_{\st}\longmapsto t q_{\st},
\qquad
q_{\bar\st}\longmapsto t^{-1}q_{\bar\st}
\]
is the internal folded little-group action.  It is the same internal scaling that appears in the ordinary momentum blob.

\subsubsection{Graph picture and fiber-product reconstruction}

\begin{figure}[h!]
\centering
\resizebox{0.8\textwidth}{!}{%
\begin{tikzpicture}
\begin{feynman}
  \draw[thick] (0,0) circle (3.2);

  \vertex[blob, minimum size=2.6cm] (C) at (0,0) {};

  \vertex[blob, minimum size=0.9cm, label=center:$L$] (L) at (-2.2,0) {};
  \vertex[blob, minimum size=0.9cm, label=center:$\bar{L}$] (R) at ( 2.2,0) {};

  \vertex[dot, label={[label distance=2pt]above left:$1$}]        (L1) at (-2.057,  2.451) {};
  \vertex[dot, label=left:$a$]                                    (L2) at (-2.900,  1.352) {};
  \vertex[dot, label=left:$a+1$]                                  (L3) at (-3.188,  0.37) {};
  \vertex[dot, label=left:$b$]                                    (L4) at (-3.188, -0.37) {};
  \vertex[dot, label=left:$b+1$]                                  (L5) at (-2.900, -1.352) {};
  \vertex[dot, label={[label distance=2pt]below left:$n$}]        (L6) at (-2.057, -2.451) {};

  \vertex[dot, label={[label distance=2pt]above right:$\bar{1}$}] (R1) at ( 2.057,  2.451) {};
  \vertex[dot, label=right:$\bar{a}$]                             (R2) at ( 2.900,  1.352) {};
  \vertex[dot, label=right:$\overline{a+1}$]                      (R3) at ( 3.188,  0.37) {};
  \vertex[dot, label=right:$\bar{b}$]                             (R4) at ( 3.188, -0.37) {};
  \vertex[dot, label=right:$\overline{b+1}$]                      (R5) at ( 2.900, -1.352) {};
  \vertex[dot, label={[label distance=2pt]below right:$\bar{n}$}] (R6) at ( 2.057, -2.451) {};

  \diagram*{
    (L1) -- (C), (L2) -- (C), (L5) -- (C), (L6) -- (C),
    (R1) -- (C), (R2) -- (C), (R5) -- (C), (R6) -- (C),
    (L3) -- (L), (L4) -- (L),
    (R3) -- (R), (R4) -- (R),
  };

  \draw[red, thick] (L.east) -- (C.west);
  \draw[red, thick] (R.west) -- (C.east);

  \node at (-2.3,  1.8) {$\vdots$};
  \node at (-3.00,  0.10) {$\vdots$};
  \node at (-2.3, -1.8) {$\vdots$};
  \node at ( 2.3,  1.8) {$\vdots$};
  \node at ( 3.00,  0.10) {$\vdots$};
  \node at ( 2.3, -1.8) {$\vdots$};
\end{feynman}
\end{tikzpicture}
}
\caption{Half-interval boundaries. Here $\bar{L}$ means the blob obtained from $L$ by reflection and color swap.}
\label{fig:diagram2}
\end{figure}
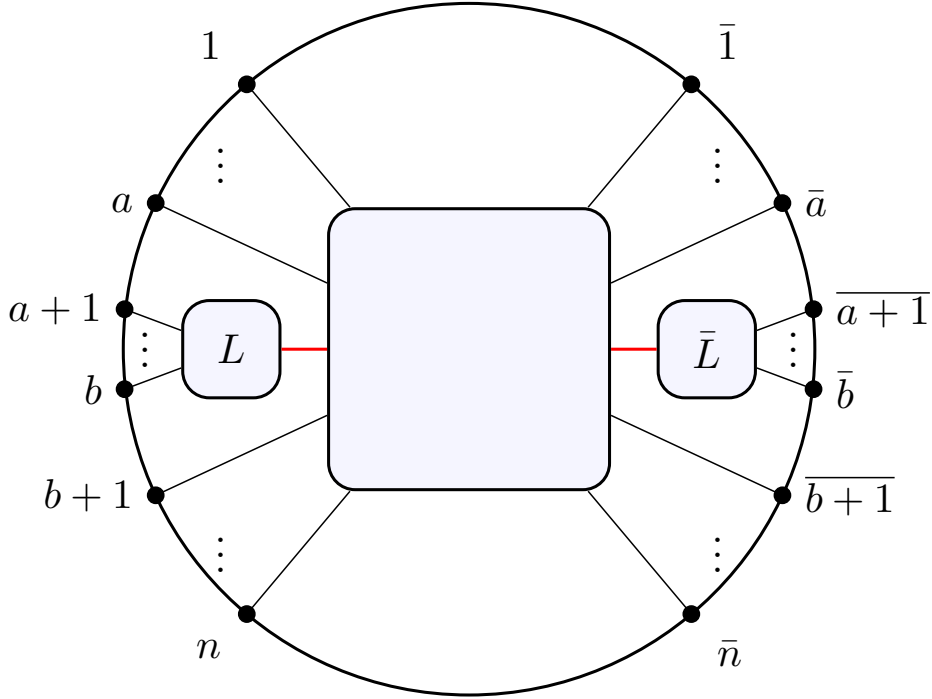

At graph level, one cuts the two reflected attachments between the ordinary blobs and the symmetric remainder.  The two cut ends become a new reflected boundary pair
\[
\st,\bar\st.
\]
The middle graph remains invariant under reflection together with color reversal, and therefore represents a cell of
\[
\LG_{\geq}^{\refl}(h,2h),
\qquad
h=n-m+1.
\]
The two outer pieces are ordinary plabic graphs.  They are exchanged by Karpman reflection and color reversal. The schematic picture is given in \cref{fig:diagram2}.

Let
\[
A:=L^{\mathrm{ext}}
\subset \mathbb R^{I\sqcup\{\st\}}
\]
be the ordinary momentum domain plane, and let
\(
A^{\refl,\vee}
\)
denote its reflected color-dual partner on
\(
\{\bar\st\}\sqcup\bar I.
\)
The full domain point is reconstructed by the double fiber product
\[
C
=
A\star_{\st}H^{\mathrm{ext}}\star_{\bar\st}A^{\refl,\vee}.
\]
Equivalently, \(C\) consists of vectors
\[
(x_I,x_K,x_{\bar I})
\]
for which there exist interface coordinates \(t,\bar t\in\mathbb R\) such that
\[
(x_I,t)\in A,
\]
\[
(t,x_K,\bar t)\in H^{\mathrm{ext}},
\]
and
\[
(\bar t,x_{\bar I})\in A^{\refl,\vee}.
\]
The rank count is
\[
(r+1)+h+(s+1)-2=n,
\]
and the ambient-size count is
\[
(m+1)+2h+(m+1)-4=2n.
\]
Thus the glued object lies in \(\Gr(n,2n)\).  The construction of the reflected dual blob and the Lagrangian condition on \(H^{\mathrm{ext}}\) ensure that the glued space is maximal isotropic, hence
\[
C\in\LG^{\refl}(n,2n).
\]
Kinematically, the ordinary blob contains the internal massless particle
\[
p_{\st}=-P_I.
\]
The reflected blob contains the reflected internal particle
\[
p_{\bar\st}=-p_{\st}^{\mathsf T}.
\]
The middle reflected Lagrangian core contains both internal labels and the remaining external labels \(K\).  Thus the boundary has the schematic form
\[
\mathcal M_{m+1,r+1}
\ \underset{\st}{\times}\
\Lampl_{n-m+1}^{\refl}
\ \underset{\bar\st}{\times}\
\left(\mathcal M_{m+1,s+1}\right)^{\refl,\vee},
\]
where the superscript \(\refl,\vee\) denotes reflected color-duality.

\subsubsection{Dimension count}

The local pair
\[
(L,\widetilde L),
\qquad
\dim L=r,
\qquad
\dim\widetilde L=s,
\qquad
L\perp\widetilde L,
\]
has dimension
\[
\dim\mathcal P_{r,s}
=
\dim\Gr(r,m)+\dim\Gr(s,s+1)
=
r(s+1)+s
=
(r+1)(s+1)-1.
\]
Equivalently, this is the dimension of the top cell of \(\Gr(r+1,m+1)\), or by parity of \(\Gr(s+1,m+1)\), minus the one-dimensional internal little-group scaling.

Adding the smaller reflected Lagrangian core gives the domain-cell dimension
\[
d_{I,r}
=
\frac{h(h+1)}2+(r+1)(s+1)-1,
\]
where
\[
h=n-m+1,
\qquad
s=m-1-r.
\]
Thus the codimension inside \(\LG^{\refl}(n,2n)\) is
\[
\operatorname{codim}
=
\frac{n(n+1)}2-d_{I,r}
=
h(m-1)+\binom r2+\binom s2.
\]

\begin{remark}
    The endpoint case \(I=[1,n]\) is included in the same description as above, in a slightly degenerate manner: here 
\(h=1\), so the smaller reflected core is \(\LG_{\geq}^{\refl}(1,2)\).
\end{remark}
\begin{remark}\label{rmk:no_r=0_codim_1}
    There is a small endpoint convention in the sector labels.  In the ordinary
momentum-amplituhedron factor over an interval \(I=[a,b]\), \(m=|I|\), the
sector \(r\) corresponds to
\[
C_{\mathrm{mom}}\in\Gr_{\geq}(r+1,m+1).
\]
The endpoint \(r=0\) forces
\[
\lambda^{\mathrm{ext}}\subset C_{\mathrm{mom}}
\]
to lie in a one-dimensional space, and hence all angle brackets on
\(I\cup\{\st\}\) vanish.  If \(m>2\), this includes at least two independent
collinear equations, for instance
\[
\langle a,a+1\rangle=\langle a+1,a+2\rangle=0.
\]
Similarly, the endpoint \(s=0\), equivalently \(r=m-1\), forces
\[
\widetilde\lambda^{\mathrm{ext}}\subset C_{\mathrm{mom}}^\perp
\]
to lie in a one-dimensional space, and hence imposes at least two square-bracket
collinear equations when \(m>2\).  Thus the endpoints \(r=0\) and \(s=0\) are
not generic codimension-one factorization sectors for intervals of length
larger than two.  For \(m=2\), however, they are exactly the two collinear
components of the reducible two-particle Mandelstam
\[
s_{[a,a+1]}=\langle a,a+1\rangle[a,a+1].
\]
The same argument works in the Lagrangian case as well, showing that for interval length greater than $2$ the $r=0$ or $s=0$ blobs do not yield codimension $1$ boundary strata of the amplituhedron.
\end{remark}

\subsection{Boundary-cell classification}
\label{subsec:reflected-boundary-cell-classification}
We summarize the preceding discussion in the following conjecture.
\begin{conjecture}
    For every positive external data $W,$ the codimension $1$ boundaries of $\Lampl_{n}^\refl(W)$ are given by the following list.
\begin{itemize}
\item For the reduced symmetric boundary
\[
\mu_r=0,
\qquad
1\leq r\leq n-1,
\]
the maximal cell is the block Lagrangian locus
\[
\LG_{\geq}^{\refl}(r,2r)
\times
\LG_{\geq}^{\refl}(n-r,2n-2r),
\]
supported on the reflection-stable decomposition
\[
[2n]=J_r\sqcup J_r^c.
\]

\item For a half-contained interval
\[
I=[a,b]\subseteq[1,n],
\qquad
m=|I|,
\]
the maximal cells over
\[
s_I=0
\]
are indexed by
\[
r+s=m-1,\qquad r,s\geq0.
\]
If $m>2$ then $r,s\geq 1.$ They have the fiber-product form
\[
C
=
A\star_{\st}H^{\mathrm{ext}}\star_{\bar\st}A^{\refl,\vee},
\]
where
\[
A\in\Gr_{\geq}(r+1,m+1),
\qquad
H^{\mathrm{ext}}
\in
\LG_{\geq}^{\refl}(n-m+1,2(n-m+1)),
\]
and the right ordinary blob is the reflected color-dual of the left one.
\end{itemize}
Modulo reflection and complement, these are the maximal reflected boundary cells associated with interval Mandelstam functions.  In particular, asymmetric axis-crossing Mandelstams do not produce new codimension-one boundaries. 
\end{conjecture}
We have verified these claims experimentally up to $n=5,$ and under the strengthened positivity assumptions discussed below, the proof of these claims will appear in \cite{MS}.
Schematically the boundary classification is given in \cref{fig:relf_BCFW}

\newcommand{\diagramZeroA}{%
  \ensuremath{\vcenter{\hbox{\resizebox{!}{0.8cm}{%
    \begin{tikzpicture}
      \begin{feynman}
          \draw[thick] (0,0) circle (3.2);
          \vertex[blob, minimum size=2.6cm] (C) at (0,0) {};

          \vertex[dot, label={[label distance=2pt]above left:$1$}]        (L1) at (-2.057,  2.451) {};
          \vertex[dot, label=left:$ $]                                    (L2) at (-2.900,  1.352) {};
          \vertex[dot, label=left:$ $]                                    (L3) at (-3.188,  0.37) {};
          \vertex[dot, label=left:$ $]                                    (L4) at (-3.188, -0.37) {};
          \vertex[dot, label=left:$ $]                                    (L5) at (-2.900, -1.352) {};
          \vertex[dot, label={[label distance=1pt]below left:$ $}]        (L6) at (-2.057, -2.451) {};

          \vertex[dot, label={[label distance=1pt]above right:$\bar{1}$}] (R1) at ( 2.057,  2.451) {};
          \vertex[dot, label=right:$ $]                                   (R2) at ( 2.900,  1.352) {};
          \vertex[dot, label=right:$ $]                                   (R3) at ( 3.188,  0.37) {};
          \vertex[dot, label=right:$ $]                                   (R4) at ( 3.188, -0.37) {};
          \vertex[dot, label=right:$ $]                                   (R5) at ( 2.900, -1.352) {};
          \vertex[dot, label={[label distance=2pt]below right:$ $}]       (R6) at ( 2.057, -2.451) {};

          \diagram*{
            (L1) -- (C), (L2) -- (C), (L5) -- (C), (L6) -- (C),
            (R1) -- (C), (R2) -- (C), (R5) -- (C), (R6) -- (C),
            (L3) -- (C), (L4) -- (C),
            (R3) -- (C), (R4) -- (C),
          };

          \node at (-3.00,  0.10) {$\vdots$};
          \node at ( 3.00,  0.10) {$\vdots$};
      \end{feynman}
    \end{tikzpicture}%
  }}}}%
}

\newcommand{\diagramOne}{%
  \ensuremath{\vcenter{\hbox{\resizebox{!}{0.8cm}{%
    \begin{tikzpicture}
      \begin{feynman}
          \draw[thick] (0,0) circle (3.2);
          \vertex[blob, minimum size=2cm] (T) at (0,1.2) {};
          \vertex[blob, minimum size=2cm] (B) at (0,-1.6) {};

          \vertex[dot, label={[label distance=1pt]above left:$1$}] (L1) at (-1.997, 2.50) {};
          \vertex[dot, label=left:$2$] (L2) at (-2.8, 1.6) {};
          \vertex[dot, label=left:$r$] (L3) at (-3.190, 0.25) {};
          \vertex[dot, label=left:$r+1$] (L4) at (-2.946, -1.25) {};
          \vertex[dot, label=left:$n$] (L5) at (-1.997, -2.50) {};

          \vertex[dot, label={[label distance=1pt]above right:$\bar{1}$}] (R1) at (1.997, 2.50) {};
          \vertex[dot, label=right:$\bar{2}$] (R2) at (2.8, 1.6) {};
          \vertex[dot, label=right:$\bar{r}$] (R3) at (3.190, 0.25) {};
          \vertex[dot, label=right:$\overline{r+1}$] (R4) at (2.946, -1.25) {};
          \vertex[dot, label=right:$\bar{n}$] (R5) at (1.997, -2.50) {};

          \diagram*{
            (L1) -- (T), (L2) -- (T), (L3) -- (T),
            (L4) -- (B), (L5) -- (B),
            (R1) -- (T), (R2) -- (T), (R3) -- (T),
            (R4) -- (B), (R5) -- (B),
          };

          \node at (-2.80, 0.95) {$\vdots$};
          \node at (-2.2, -1.8) {$\vdots$};
          \node at (2.80, 0.95) {$\vdots$};
          \node at (2.2, -1.8) {$\vdots$};
      \end{feynman}
    \end{tikzpicture}%
  }}}}%
}

\newcommand{\diagramTwo}{%
  \ensuremath{\vcenter{\hbox{\resizebox{!}{0.8cm}{%
    \begin{tikzpicture}
      \begin{feynman}
          \draw[thick] (0,0) circle (3.2);
          \vertex[blob, minimum size=2.6cm] (C) at (0,0) {};
          \vertex[blob, minimum size=0.9cm, label=center:$L$] (L) at (-2.2,0) {};
          \vertex[blob, minimum size=0.9cm, label=center:$\bar{L}$] (R) at ( 2.2,0) {};

          \vertex[dot, label={[label distance=2pt]above left:$1$}]        (L1) at (-2.057,  2.451) {};
          \vertex[dot, label=left:$a$]                                    (L2) at (-2.900,  1.352) {};
          \vertex[dot, label=left:$a+1$]                                  (L3) at (-3.188,  0.37) {};
          \vertex[dot, label=left:$b$]                                    (L4) at (-3.188, -0.37) {};
          \vertex[dot, label=left:$b+1$]                                  (L5) at (-2.900, -1.352) {};
          \vertex[dot, label={[label distance=2pt]below left:$n$}]        (L6) at (-2.057, -2.451) {};

          \vertex[dot, label={[label distance=2pt]above right:$\bar{1}$}] (R1) at ( 2.057,  2.451) {};
          \vertex[dot, label=right:$\bar{a}$]                             (R2) at ( 2.900,  1.352) {};
          \vertex[dot, label=right:$\overline{a+1}$]                      (R3) at ( 3.188,  0.37) {};
          \vertex[dot, label=right:$\bar{b}$]                             (R4) at ( 3.188, -0.37) {};
          \vertex[dot, label=right:$\overline{b+1}$]                      (R5) at ( 2.900, -1.352) {};
          \vertex[dot, label={[label distance=2pt]below right:$\bar{n}$}] (R6) at ( 2.057, -2.451) {};

          \diagram*{
            (L1) -- (C), (L2) -- (C), (L5) -- (C), (L6) -- (C),
            (R1) -- (C), (R2) -- (C), (R5) -- (C), (R6) -- (C),
            (L3) -- (L), (L4) -- (L),
            (R3) -- (R), (R4) -- (R),
          };

          \draw[red, thick] (L.east) -- (C.west);
          \draw[red, thick] (R.west) -- (C.east);

          \node at (-2.3,  1.8) {$\vdots$};
          \node at (-3.00,  0.10) {$\vdots$};
          \node at (-2.3, -1.8) {$\vdots$};
          \node at ( 2.3,  1.8) {$\vdots$};
          \node at ( 3.00,  0.10) {$\vdots$};
          \node at ( 2.3, -1.8) {$\vdots$};
      \end{feynman}
    \end{tikzpicture}%
  }}}}%
}

\newcommand{\diagramZeroB}{%
  \ensuremath{\vcenter{\hbox{\resizebox{!}{0.8cm}{%
    \begin{tikzpicture}
      \begin{feynman}
          \draw[thick] (0,0) circle (3.2);
          \vertex[blob, minimum size=2.6cm] (C) at (0,0) {};

          \vertex[dot, label={[label distance=2pt]above left:$1$}]        (L1) at (-2.057,  2.451) {};
          \vertex[dot, label=left:$ $]                                    (L2) at (-2.900,  1.352) {};
          \vertex[dot, label=left:$ $]                                    (L3) at (-3.188,  0.37) {};
          \vertex[dot, label=left:$ $]                                    (L4) at (-3.188, -0.37) {};
          \vertex[dot, label=left:$ $]                                    (L5) at (-2.900, -1.352) {};
          \vertex[dot, label={[label distance=1pt]below left:$ $}]        (L6) at (-2.057, -2.451) {};

          \vertex[dot, label={[label distance=1pt]above right:$2n$}]      (R1) at ( 2.057,  2.451) {};
          \vertex[dot, label=right:$ $]                                   (R2) at ( 2.900,  1.352) {};
          \vertex[dot, label=right:$ $]                                   (R3) at ( 3.188,  0.37) {};
          \vertex[dot, label=right:$ $]                                   (R4) at ( 3.188, -0.37) {};
          \vertex[dot, label=right:$ $]                                   (R5) at ( 2.900, -1.352) {};
          \vertex[dot, label={[label distance=2pt]below right:$ $}]       (R6) at ( 2.057, -2.451) {};

          \diagram*{
            (L1) -- (C), (L2) -- (C), (L5) -- (C), (L6) -- (C),
            (R1) -- (C), (R2) -- (C), (R5) -- (C), (R6) -- (C),
            (L3) -- (C), (L4) -- (C),
            (R3) -- (C), (R4) -- (C),
          };

          \node at (-3.00,  0.10) {$\vdots$};
          \node at ( 3.00,  0.10) {$\vdots$};
      \end{feynman}
    \end{tikzpicture}%
  }}}}%
}

\newcommand{\diagramThree}{%
  \ensuremath{\vcenter{\hbox{\resizebox{!}{0.8cm}{%
    \begin{tikzpicture}
      \begin{feynman}
          \draw[thick] (0,0) circle (3.2);

          \vertex[blob, minimum size=1.5cm, label=center:{\textcolor{red}{$\rho (L)$}}] (T) at (1,1) {};
          \vertex[blob, minimum size=1.5cm, label=center:$L$] (B) at (-1,-1) {};

          \vertex[dot, label={[label distance=2pt]above left:$r$}]        (v0) at (-3.043, 0.989) {};
          \vertex[dot, label={[label distance=2pt]above left:$r-1$}]      (v1) at (-1.882, 2.589) {};
          \vertex[dot, label={[label distance=2pt]above:$r-2$}]           (v2) at (0.000, 3.200) {};
          \vertex[dot, label={[label distance=2pt]right:$r+n+1$}]         (v4) at (3.043, 0.989) {};
          \vertex[dot, label={[label distance=2pt]right:$r+n$}]           (v5) at (3.043, -0.989) {};
          \vertex[dot, label={[label distance=3.5pt]below:$r+n-1$}]       (v6) at (1.882, -2.589) {};
          \vertex[dot, label={[label distance=2pt]below:$r+n-2$}]         (v7) at (0.000, -3.200) {};
          \vertex[dot, label={[label distance=2pt]below left:$r+1$}]      (v9) at (-3.043, -0.989) {};

          \diagram*{
            (v0) -- (B), (v1) -- (T), (v2) -- (T), (v4) -- (T),
            (v5) -- (T), (v6) -- (B), (v7) -- (B), (v9) -- (B)
          };

          \draw[red, thick] (T.south west) -- (B.north east);
      \end{feynman}
      \node at (2,  2.1) {$\vdots$};
      \node at (-2,  -2) {$\vdots$};
    \end{tikzpicture}%
  }}}}%
}

\newcommand{\diagramFour}{%
  \ensuremath{\vcenter{\hbox{\resizebox{!}{0.8cm}{%
    \begin{tikzpicture}
      \begin{feynman}
          \draw[thick] (0,0) circle (3.2);
          \vertex[blob, minimum size=2.6cm] (C) at (0,0) {};
          \vertex[blob, minimum size=0.9cm, label=center:$L$] (L) at (-2.2,0) {};
          \vertex[blob, minimum size=0.9cm, label=center:{\textcolor{red}{$\rho (L)$}}] (R) at ( 2.2,0) {};

          \vertex[dot, label={[label distance=2pt]above left:$1$}]        (L1) at (-2.057,  2.451) {};
          \vertex[dot, label=left:$a-1$]                                  (L2) at (-2.900,  1.352) {};
          \vertex[dot, label=left:$a$]                                    (L3) at (-3.188,  0.37) {};
          \vertex[dot, label=left:$b$]                                    (L4) at (-3.188, -0.37) {};
          \vertex[dot, label=left:$b+1$]                                  (L5) at (-2.000, -2.5) {};

          \vertex[dot, label={[label distance=2pt]above right:$2n$}]      (R1) at ( 2.057,  2.451) {};
          \vertex[dot, label=right:$b+n+1$]                               (R2) at ( 2.900,  1.352) {};
          \vertex[dot, label=right:$b+n$]                                 (R3) at ( 3.188,  0.37) {};
          \vertex[dot, label=right:$a+n$]                                 (R4) at ( 3.188, -0.37) {};
          \vertex[dot, label=right:$a+n-1$]                               (R5) at ( 2.000, -2.5) {};

          \diagram*{
            (L1) -- (C), (L2) -- (C), (L5) -- (C), 
            (R1) -- (C), (R2) -- (C), (R5) -- (C), 
            (L3) -- (L), (L4) -- (L),
            (R3) -- (R), (R4) -- (R),
          };

          \draw[red, thick] (L.east) -- (C.west);
          \draw[red, thick] (R.west) -- (C.east);
      \end{feynman}

      \node at (-2.3,  1.8) {$\vdots$};
      \node at (-3.00,  0.10) {$\vdots$};
      \node at ( 2.3,  1.8) {$\vdots$};
      \node at ( 3.00,  0.10) {$\vdots$};
      \node at (0, -2.2) {$\dots$};
    \end{tikzpicture}%
  }}}}%
}

\begin{figure}[h!]
\centering
\scalebox{2.3}{%
  $\displaystyle
  \partial \biggl( \diagramZeroA \biggr)
  =
  \sum_{r} \diagramOne
  +
  \sum_{a,b,\operatorname{rank}(L)} \diagramTwo
  $%
}
\caption{Scheme of the BCFW boundary classification}
\label{fig:relf_BCFW}
\end{figure}

\subsection{BCFW tiling and recursion}
\label{subsec:reflected-bcfw-tiling}

We now describe the reflected BCFW tilings. We start with a single step of this recursive construction: the decomposition of $\Lampl_{n}^\refl$ into subspaces, associated to a fixed bridge type $M$ of the form \eqref{eq:right_refl_bridge},~\eqref{eq:left_refl_bridge},~\eqref{eq:refl_bridge_mid} or~\eqref{eq:refl_bridge_out}, for a fixed $i\in[n]\cup\overline{[n]}.$

Fix one of
the reflected bridges~\eqref{eq:right_refl_bridge},~\eqref{eq:left_refl_bridge}
\[
M(t)=M_i^{\refl,\bullet}(t),
\qquad
\bullet\in\{\rightarrow,\leftarrow\},
\]
or one of the reflection-axis bridges~\eqref{eq:refl_bridge_mid},~\eqref{eq:refl_bridge_out}. 

Let 
\(\mathcal B\) be a boundary cell of one of the two types above. Let $F_{\mathcal B}$ be the associated boundary functionary, by the discussion in the previous subsections, either
\(F_{\mathcal B}=\mu_r\) or \(F_{\mathcal B}=s_I\) for $I=[a,b]\subseteq[1,n]$. For a fixed bridge $M$ only the boundary cells $\mathcal{B}$ whose defining function depends nontrivially on the bridge parameter in the
bridge-removal equation of \cref{subsec:bridge-promotion-principle} will be used for the decomposition. We say that these boundary cells are \emph{hit} by $M.$ For such $\mathcal{B}$ add the bridge $M$ to obtain the domain cell 
\[
\mathcal T_{\mathcal B}
=
\mathcal B\cdot M(\mathbb R_{>0})
=
\{\,C_0M(t): C_0\in\mathcal B,\ t>0\,\}
\subset
\LG_{\geq}^{\refl}(n,2n).
\]
The corresponding image cell is
\[
\Phi_W(\mathcal T_{\mathcal B})
\subset
\Lampl_n^{\refl}(W).
\]
For a point in a bridged
cell $\mathcal{T}_{\mathcal{B}}$, write
\[
C'=C_0M(t),~C_0\in\mathcal{B},~t\in\mathbb{R}_+.
\]
Let $Y=\Phi_W(C).$ Crucially for the recursion, we can recover the bridge parameter $t$ from $Y.$ To this end, we move the bridge to the external data as in
\eqref{eq:bridge-transfer-identity}. At the spinor level this means using the
substitution \eqref{eq:bridge-spinor-substitution}. Thus we form
\[
F_{\mathcal B}^{M}(t)
:=
F_{\mathcal B}
\bigl(
\widetilde\lambda M(t)^{\mathsf T},
\lambda M(t)^{-1}
\bigr).
\]
Since $F_\mathcal{B}$ vanishes on the image of $\mathcal{B},$ we obtain the bridge-removal equation:
\[
F_{\mathcal B}^{M}(t)=0.
\]
In the reflected model, the primary boundary functionaries give affine-linear
bridge-removal equations.  To make the coefficient explicit, we demonstrate it in the case of a
right paired bridge
\[
M(t)=M_i^{\refl,\rightarrow}(t),
\qquad
1\leq i\leq n-1.
\]
Under the bridge-removal substitution
\eqref{eq:bridge-spinor-substitution}, the shifted spinors are
\[
\widetilde\lambda_i(t)=\widetilde\lambda_i+t\widetilde\lambda_{i+1},
\qquad
\widetilde\lambda_{\overline{i+1}}(t)
=
\widetilde\lambda_{\overline{i+1}}+t\widetilde\lambda_{\bar i},
\]
and
\[
\lambda_{i+1}(t)=\lambda_{i+1}-t\lambda_i,
\qquad
\lambda_{\bar i}(t)=\lambda_{\bar i}-t\lambda_{\overline{i+1}}.
\]
For the symmetric boundary functionary
\[
F_{\mathcal B}=\mu_r,
\]
only the function corresponding to the crossed cut changes.  More precisely,
\begin{equation}
\label{eq:refl-dot-mu}
\mu_r^M(t)
=
\mu_r+t\,\dot\mu_{r,i},
\qquad
\dot\mu_{r,i}
=
\begin{cases}
(-1)^{i+1}[\,i{+}1,\bar i\,],& r=i,\\
0,& r\neq i.
\end{cases}
\end{equation}
Thus the bridge hits the symmetric boundary precisely when \(r=i\), and then
\[
t_{\mu_i}
=
-\frac{\mu_i}{(-1)^{i+1}[\,i{+}1,\bar i\,]}
=
\frac{(-1)^i\mu_i}{[\,i{+}1,\bar i\,]}.
\]
For a half-contained Mandelstam boundary
\[
F_{\mathcal B}=s_I,
\qquad
I=[a,b]\subseteq[1,n],
\]
set
\[
\epsilon_I^{(i)}
:=
\mathbf 1_I(i)-\mathbf 1_I(i+1).
\]
Then
\begin{equation}
\label{eq:refl-dot-half-mandelstam}
s_I^M(t)
=
s_I+t\,\dot s_{I,i},
\qquad
\dot s_{I,i}
=
\epsilon_I^{(i)}F_{I;i},
\end{equation}
where
\begin{equation}
\label{eq:refl-F-I-i}
F_{I;i}
=
\sum_{j\in I}
\langle i,j\rangle[\,i{+}1,j\,].
\end{equation}
Using the reflected folding relation, this can also be written purely in
square brackets as
\[
F_{I;i}
=
\sum_{j\in I}
(-1)^{i+j}
[\,\bar i,\bar j\,]\,[\,i{+}1,j\,].
\]
Thus \(s_I\) depends on the bridge parameter exactly when the interval \(I\)
separates the adjacent labels \(i\) and \(i+1\).  In that case the
bridge-removal value is
\[
t_I
=
-\frac{s_I}{\epsilon_I^{(i)}F_{I;i}}.
\]
Therefore, for any hit boundary \(\mathcal B\), we may write uniformly
\[
F_{\mathcal B}^{M}(t)
=
F_{\mathcal B}+t\dot F_{\mathcal B,M},
\]
where \(\dot F_{\mathcal B,M}\) is given by
\eqref{eq:refl-dot-mu} for \(F_{\mathcal B}=\mu_r\), and by
\eqref{eq:refl-dot-half-mandelstam} for \(F_{\mathcal B}=s_I\).  When
\[
\dot F_{\mathcal B,M}\neq0,
\]
the bridge hits the boundary and
\begin{equation}\label{eq:t_B}
t_{\mathcal B}
=
-\frac{F_{\mathcal B}}{\dot F_{\mathcal B,M}},
\end{equation}
while if $
\dot F_{\mathcal B,M}=0,$ the boundary is not hit and will not contribute to the BCFW decomposition associated to $M.$

For a paired bridge away from the reflection axes, exactly one reduced
symmetric function changes: the one corresponding to the cut crossed by the
bridge. Thus a bridge crossing the cut \(J_i\mid J_i^c\) hits the symmetric
boundary
\[
\mu_i=0.
\]
A reflection-axis bridge hits no symmetric boundary, since every symmetric
interval contains either both labels of the axis pair or neither of them.

For half-contained Mandelstams, a bridge hits precisely those intervals
\[
I=[a,b]\subseteq[1,n]
\]
which separate the two adjacent labels in the ordinary part of the bridge. For
two-particle intervals one first factors
\[
s_{[a,a+1]}
=
[a,a+1]\,[\overline{a+1},\bar a].
\]
Only the irreducible collinear component which depends on the bridge parameter
produces a finite BCFW root; the shift-invariant component does not contribute
for this bridge.

For each hit boundary \(\mathcal B\), attaching the bridge adds one positive
parameter. Since the boundary image has dimension \(2n-2\), the bridged image
has the expected full dimension
\[
2n-1
=
\dim \Lampl_n^{\refl}(W).
\]

\begin{conjecture}[Reflected BCFW tiling]
\label{conj:reflected-bcfw-tiling}
Fix a reflected positive bridge \(M(t)\). Assume that the external datum \(W\)
is positive, and choose the boundary components
hit by \(M(t)\) as above. Then the images of the cells
\[
\mathcal T_{\mathcal B}
=
\mathcal B\cdot M(\mathbb R_{>0})
\]
tile the reflected Lagrangian amplituhedron:
\[
\Lampl_n^{\refl}(W)
=
\bigcup_{\mathcal B\in\operatorname{Hit}(M)}
\Phi_W(\mathcal T_{\mathcal B}).
\]
Equivalently, the BCFW cells, calculated by iterating the BCFW procedure, tile the amplituhedron.
\end{conjecture}

The proof in the case that $W$ is strongly positive, in the sense below, will appear in \cite{MS}. The statement
is supported by the exact computations summarized in
\cref{app:computational-evidence}. 
The diagrammatic equations below describe the geometric one-step BCFW recursion in the cases of bridges $M_1^{\refl,\rightarrow}$ and $M_{\operatorname{out}}^{\refl,\leftarrow}$
\newcommand{\diagramreflzero}{%
  \ensuremath{\vcenter{\hbox{\resizebox{!}{0.8cm}{%
    \begin{tikzpicture}
      \begin{feynman}
          \draw[thick] (0,0) circle (3.2);
          \vertex[blob, minimum size=2.6cm] (C) at (0,0) {};

          \vertex[dot, label={[label distance=2pt]above left:$1$}]        (L1) at (-2.057,  2.451) {};
          \vertex[dot, label=left:$ $]                                    (L2) at (-2.900,  1.352) {};
          \vertex[dot, label=left:$ $]                                    (L3) at (-3.188,  0.37) {};
          \vertex[dot, label=left:$ $]                                    (L4) at (-3.188, -0.37) {};
          \vertex[dot, label=left:$ $]                                    (L5) at (-2.900, -1.352) {};
          \vertex[dot, label={[label distance=1pt]below left:$n$}]        (L6) at (-2.057, -2.451) {};

          \vertex[dot, label={[label distance=1pt]above right:$\bar{1}$}] (R1) at ( 2.057,  2.451) {};
          \vertex[dot, label=right:$ $]                                   (R2) at ( 2.900,  1.352) {};
          \vertex[dot, label=right:$ $]                                   (R3) at ( 3.188,  0.37) {};
          \vertex[dot, label=right:$ $]                                   (R4) at ( 3.188, -0.37) {};
          \vertex[dot, label=right:$ $]                                   (R5) at ( 2.900, -1.352) {};
          \vertex[dot, label={[label distance=2pt]below right:$\bar{n}$}]       (R6) at ( 2.057, -2.451) {};

          \diagram*{
            (L1) -- (C), (L2) -- (C), (L5) -- (C), (L6) -- (C),
            (R1) -- (C), (R2) -- (C), (R5) -- (C), (R6) -- (C),
            (L3) -- (C), (L4) -- (C),
            (R3) -- (C), (R4) -- (C),
          };

          \node at (-3.00,  0.10) {$\vdots$};
          \node at ( 3.00,  0.10) {$\vdots$};
      \end{feynman}
    \end{tikzpicture}%
  }}}}%
}

\newcommand{\diagramreflone}{%
  \ensuremath{\vcenter{\hbox{\resizebox{!}{0.8cm}{%
    \begin{tikzpicture}
      \begin{feynman}
          \draw[thick] (0,0) circle (3.2);
          \vertex[blob, minimum size=2.6cm] (C) at (0,0) {};

          \vertex[dot, label={[label distance=2pt]above left:$1$}]        (L1) at (-2.057,  2.451) {};
          \vertex[dot, label=left:$2$]                                    (L2) at (-2.900,  1.352) {};
          \vertex[dot, label=left:$ $]                                    (L3) at (-3.188,  0.37) {};
          \vertex[dot, label=left:$ $]                                    (L4) at (-3.188, -0.37) {};
          \vertex[dot, label=left:$ $]                                    (L5) at (-2.900, -1.352) {};
          \vertex[dot, label={[label distance=1pt]below left:$n$}]        (L6) at (-2.057, -2.451) {};

          \vertex[dot, label={[label distance=1pt]above right:$\bar{1}$}] (R1) at ( 2.057,  2.451) {};
          \vertex[dot, label=right:$\bar{2}$]                                   (R2) at ( 2.900,  1.352) {};
          \vertex[dot, label=right:$ $]                                   (R3) at ( 3.188,  0.37) {};
          \vertex[dot, label=right:$ $]                                   (R4) at ( 3.188, -0.37) {};
          \vertex[dot, label=right:$ $]                                   (R5) at ( 2.900, -1.352) {};
          \vertex[dot, label={[label distance=2pt]below right:$\bar{n}$}]       (R6) at ( 2.057, -2.451) {};

  \vertex[dot, minimum size=8pt, fill=white, draw=black] (B1) at (-1.2,1.5) {};
  \vertex[dot, minimum size=8pt, fill=black, draw=black] (B2) at (-1.7,0.8) {};
  \vertex[dot, minimum size=8pt, fill=white, draw=black] (B3) at (1.7,0.8) {};
  \vertex[dot, minimum size=8pt, fill=black, draw=black] (B4) at (1.2,1.5) {};

          \diagram*{
            (L1) -- (B1), (L2) -- (B2), (L5) -- (C), (L6) -- (C),
            (R1) -- (B4), (R2) -- (B3), (R5) -- (C), (R6) -- (C),
            (L3) -- (C), (L4) -- (C),
            (R3) -- (C), (R4) -- (C),
            (B1) -- (C), (B2) -- (C), (B3) -- (C), (B4) -- (C),
            (B1) -- (B2), (B3) -- (B4),
          };

          \node at (-3.00,  0.10) {$\vdots$};
          \node at ( 3.00,  0.10) {$\vdots$};
      \end{feynman}
    \end{tikzpicture}%
  }}}}%
}

\newcommand{\diagramrefltwo}{%
  \ensuremath{\vcenter{\hbox{\resizebox{!}{0.8cm}{%
    \begin{tikzpicture}
\begin{feynman}
  \draw[thick] (0,0) circle (3.2);

  \vertex[blob, minimum size=2.6cm] (C) at (0,-0.75) {};

  \vertex[blob, minimum size=0.9cm, label=center:$L$] (L) at (-2,0.5) {};
  \vertex[blob, minimum size=0.9cm, label=center:$\overline{L}$] (R) at (2,0.5) {};

  \vertex[dot, label={[label distance=2pt]above left:$1$}]        (L1) at (-1.8,  2.6) {};
  \vertex[dot, label=left:$2$]                                  (L3) at (-2.65,  1.8) {};
  \vertex[dot, label=left:$b$]                                    (L4) at (-3.19, -0.4) {};
  \vertex[dot, label=left:$b+1$]                                  (L5) at (-2.900, -1.352) {};

  \vertex[dot, label={[label distance=2pt]above right:$\overline{1}$}] (R1) at (1.8,  2.6) {};
  \vertex[dot, label=right:$\overline{2}$]                      (R3) at ( 2.65,  1.8) {};
  \vertex[dot, label=right:$\overline{b}$]                             (R4) at ( 3.188, -0.37) {};
  \vertex[dot, label=right:$\overline{b+1}$]                      (R5) at ( 2.900, -1.352) {};

  \vertex[dot, minimum size=8pt, fill=white, draw=black] (B1) at (-1.45,2.0) {};
  \vertex[dot, minimum size=8pt, fill=black, draw=black] (B2) at (-2.4,1.3) {};
  \vertex[dot, minimum size=8pt, fill=white, draw=black] (B3) at (2.4,1.3) {};
  \vertex[dot, minimum size=8pt, fill=black, draw=black] (B4) at (1.45,2.0) {};

  \diagram*{
    (L1) -- (B1), (B1) -- (C), (L5) -- (C),
    (R1) -- (B4), (B4) -- (C), (R5) -- (C),
    (L3) -- (B2), (B2) -- (L), (L4) -- (L),
    (R3) -- (B3), (B3) -- (R), (R4) -- (R),
    (B1) -- (B2), (B3) -- (B4),
  };

  \draw[black, thick] (L.south east) -- (C.west);
  \draw[black, thick] (R.south west) -- (C.east);

  \node at (-2.7,  0.50) {$\vdots$};
  \node at (0, -2.5) {$\dots$};
  \node at ( 2.7,  0.50) {$\vdots$};
\end{feynman}
\end{tikzpicture}%
  }}}}%
}

\begin{equation}\label{eq:bcfw_right_refl}
\scalebox{2.3}{%
  $\displaystyle
    \diagramreflzero
  =
  \diagramreflone
  +
  \sum_{b,\operatorname{rank}(L)} \diagramrefltwo
  $%
}
\end{equation}


\newcommand{\diagramreflalt}{%
  \ensuremath{\vcenter{\hbox{\resizebox{!}{0.8cm}{%
    \begin{tikzpicture}
\begin{feynman}
  \draw[thick] (0,0) circle (3.2);

  \vertex[blob, minimum size=2.6cm] (C) at (0,-0.75) {};

  \vertex[blob, minimum size=0.9cm, label=center:$L$] (L) at (-2,0.5) {};
  \vertex[blob, minimum size=0.9cm, label=center:$\overline{L}$] (R) at (2,0.5) {};

  \vertex[dot, label={[label distance=2pt]above left:$1$}]        (L1) at (-1.8,  2.6) {};
  \vertex[dot, label=left:$2$]                                  (L3) at (-2.65,  1.8) {};
  \vertex[dot, label=left:$b$]                                    (L4) at (-3.19, -0.4) {};
  \vertex[dot, label=left:$b+1$]                                  (L5) at (-2.900, -1.352) {};

  \vertex[dot, label={[label distance=2pt]above right:$\overline{1}$}] (R1) at (1.8,  2.6) {};
  \vertex[dot, label=right:$\overline{2}$]                      (R3) at ( 2.65,  1.8) {};
  \vertex[dot, label=right:$\overline{b}$]                             (R4) at ( 3.188, -0.37) {};
  \vertex[dot, label=right:$\overline{b+1}$]                      (R5) at ( 2.900, -1.352) {};

  \vertex[dot, minimum size=8pt, fill=white, draw=black] (B1) at (-1.45,2.0) {};
  \vertex[dot, minimum size=8pt, fill=black, draw=black] (B4) at (1.45,2.0) {};

  \diagram*{
    (L1) -- (B1), (B1) -- (L), (L5) -- (C),
    (R1) -- (B4), (B4) -- (R), (R5) -- (C),
    (L3) -- (L), (L4) -- (L),
    (R3) -- (R), (R4) -- (R),
    (B1) -- (B4)
  };

  \draw[black, thick] (L.south east) -- (C.west);
  \draw[black, thick] (R.south west) -- (C.east);

  \node at (-2.7,  0.50) {$\vdots$};
  \node at (0, -2.5) {$\dots$};
  \node at ( 2.7,  0.50) {$\vdots$};
\end{feynman}
\end{tikzpicture}%
  }}}}%
}

\begin{equation}\label{eq:bcfw_out_diag}
\scalebox{2.3}{%
  $\displaystyle
    \diagramreflzero
  =
  \sum_{b,\operatorname{rank}(L)} \diagramreflalt
  $%
}
\end{equation}

\subsubsection{Recursive description of the cells}
A hit symmetric boundary contributes a subspace of the form
\[
\Phi_W(\mathcal T_{\mu_r})
=
\Phi_W\left(
\left[\LG_{\geq}^{\refl}(r,2r)
\times
\LG_{\geq}^{\refl}(n-r,2n-2r)\right]
\cdot M(\mathbb R_{>0})
\right),
\]
where the bridge crosses the cut \(J_r\mid J_r^c\).

A hit half-contained interval
\[
I=[a,b]\subseteq[1,n],
\qquad
m=|I|,
\]
contributes one subspace for each
\[
r+s=m-1,\qquad r,s\geq 1,
\]
and if $m=2,~r,s\geq 0$ (see \cref{rmk:no_r=0_codim_1}).
Using the notation of \cref{subsec:reflected-half-interval-boundaries}, the
boundary cell has the fiber-product form
\[
\mathcal B_{I,r}
=
A\star_{\st}H^{\mathrm{ext}}\star_{\bar\st}A^{\refl,\vee},
\]
where, in the standard momentum-amplituhedron convention,
\[
A\in\Gr_{\geq}(r+1,m+1),
\qquad
H^{\mathrm{ext}}
\in
\LG_{\geq}^{\refl}(n-m+1,2(n-m+1)).
\]
The corresponding subspace is
\[
\Phi_W(\mathcal T_{I,r})
=
\Phi_W(\mathcal B_{I,r}\cdot M(\mathbb R_{>0})).
\]
In both cases one can then factorize the smaller reflected or momentum amplituhedra, via the same BCFW procedure, to finally obtain a collection of $2n-1$-dimensional domain cells that form a tiling the reflected Lagrangian amplituhedron.


\subsubsection{Canonical forms}
We finally state the expected recursion for canonical forms. 
Let
\[
\Omega_n^{\refl}
\]
denote the conjectural canonical form of \(\Lampl_n^{\refl}(W)\).  Fix a hit boundary
component \(\mathcal B\), and let \(M(t)\) be the bridge used in the
corresponding BCFW term.  The lower forms appearing below are not evaluated on
the original external datum \(W\), but on the shifted external datum
\begin{equation}
\label{eq:shifted-external-data-for-forms}
W_{\mathcal B}(t):=M(t)W.
\end{equation}
Equivalently, in the spinor-helicity \(B\)-picture, they are evaluated on the
shifted spinors
\begin{equation}
\label{eq:shifted-spinors-for-forms}
\widetilde\lambda_{\mathcal B}(t)
=
\widetilde\lambda M(t)^{\mathsf T},
\qquad
\lambda_{\mathcal B}(t)
=
\lambda M(t)^{-1}.
\end{equation}
This is the same shift as in \eqref{eq:bridge-spinor-substitution}.

On a symmetric boundary
\[
\mu_r=0,
\]
the shifted boundary kinematics splits into the two reflected subsystems
supported on
\[
J_r
\qquad\text{and}\qquad
J_r^c.
\]
The boundary form  therefore factorizes as 
\[
\Omega_r^{\refl}
\bigl(W_{\mathcal B}(t)|_{J_r}\bigr)
\wedge
\Omega_{n-r}^{\refl}
\bigl(W_{\mathcal B}(t)|_{J_r^c}\bigr).
\]
When the shifted external data are clear from context, we abbreviate this as
\[
\Omega_r^{\refl}\wedge\Omega_{n-r}^{\refl}.
\]
Indeed,
\[
(2r-1)+(2(n-r)-1)=2n-2.
\]
On a half-interval boundary
\[
I=[a,b]\subseteq[1,n],
\qquad
m=|I|,
\]
the component $\mathcal{B}_{I,r}$ indexed by \(r+s=m-1\) contains an ordinary momentum-amplituhedron
factor
\[
\mathcal M_{m+1,r+1}
\]
and a smaller reflected Lagrangian factor
\[
\Lampl_{n-m+1}^{\refl}.
\]
At the shifted kinematic point
\[
\bigl(\widetilde\lambda_{\mathcal B}(t),
\lambda_{\mathcal B}(t)\bigr),
\]
the equation \(s_I=0\) determines an internal massless particle
\[
P_I(t)
=
-\lambda_{\st}(t)\widetilde\lambda_{\st}(t)^{\mathsf T},
\]
unique up to the internal little group.  The reflected internal particle is
then fixed by
\[
p_{\bar\st}(t)=-p_{\st}(t)^{\mathsf T}.
\]
After quotienting by the internal little group, we denote the shifted boundary
form by
\[
\Omega_{I,r}^{\mathrm{fac}}(t)
:=
\Omega_{m+1,r+1}^{\mathrm{mom}}(I\cup\{\st\};t)
\mathop{\star}_{\st}
\Omega_{n-m+1}^{\refl}(K\cup\{\st,\bar\st\};t).
\]
Here the argument \(t\) means that both lower forms are evaluated on the
shifted data from \eqref{eq:shifted-external-data-for-forms}, or equivalently
on the shifted spinors from \eqref{eq:shifted-spinors-for-forms}.  When no
confusion is likely, we abbreviate this as
\[
\Omega_{I,r}^{\mathrm{fac}}
:=
\Omega_{m+1,r+1}^{\mathrm{mom}}
\mathop{\star}_{\st}
\Omega_{n-m+1}^{\refl}.
\]
Equivalently, let
\[
\widehat{\mathcal B}_{I,r}(t)
\]
be the total space of the fiber bundle over the shifted half-interval boundary on which the internal
spinors
\[
\lambda_{\st}(t),\qquad \widetilde\lambda_{\st}(t)
\]
are chosen, rather than quotiented by the internal little group
\[
\lambda_{\st}(t)\longmapsto z_{\st}\lambda_{\st}(t),
\qquad
\widetilde\lambda_{\st}(t)\longmapsto
z_{\st}^{-1}\widetilde\lambda_{\st}(t).
\]
The reflected internal spinors at \(\bar\st\) are then fixed by the folded
relation.  We denote by
\[
\pi:\widehat{\mathcal B}_{I,r}(t)\longrightarrow \mathcal B_{I,r}(t)
\]
the quotient map which forgets this internal little-group choice.  On this
cover one has
\[
\Omega_{m+1,r+1}^{\mathrm{mom}}(I\cup\{\st\};t)
\wedge
\Omega_{n-m+1}^{\refl}(K\cup\{\st,\bar\st\};t)
=
d\log z_{\st}\wedge\pi^*\Omega_{I,r}^{\mathrm{fac}}(t),
\]
up to orientation.

Let
\[
\Psi_{\mathcal B}:
\mathcal B\times\mathbb R_{>0}
\longrightarrow
\Lampl_n^{\refl}(W)
\]
be the map obtained by attaching the bridge \(M(t)\) and applying \(\Phi_W\):
\[
\Psi_{\mathcal B}(C_0,t)=\Phi_W(C_0M(t)).
\]
This discussion leads us to the next main conjecture of the paper:
\begin{conjecture}
The reflected Lagrangian amplituhedron is a positive geometry in the sense of \cite{ArkaniHamedBaiLam2017PositiveGeometries}, and its canonical form is given by the formula
\begin{equation}\label{eq:BCFW_rec_refl}
\Omega_n^{\refl}
=
\sum_{\mathcal B\in\operatorname{Hit}(M)}
\,
(\Psi_{\mathcal B})_*
\left(
 d\log t_{\mathcal B}\wedge\Omega_{\mathcal B}(t_{\mathcal B})
\right),
\end{equation}
where \(t_{\mathcal B}\) is given in \eqref{eq:t_B}.  The shifted boundary form
\(\Omega_{\mathcal B}(t_{\mathcal B})\) is
\[
\Omega_{\mathcal B}(t_{\mathcal B})
=
\begin{cases}
\Omega_r^{\refl}
\bigl(W_{\mathcal B}(t_{\mathcal B})|_{J_r}\bigr)
\wedge
\Omega_{n-r}^{\refl}
\bigl(W_{\mathcal B}(t_{\mathcal B})|_{J_r^c}\bigr),
&\text{for a symmetric boundary},\\[2mm]
\Omega_{m+1,r+1}^{\mathrm{mom}}(I\cup\{\st\};t_{\mathcal B})
\mathop{\star}_{\st}
\Omega_{n-m+1}^{\refl}(K\cup\{\st,\bar\st\};t_{\mathcal B}),
&\text{for a half-interval boundary}.
\end{cases}
\]
\end{conjecture}
Note that the smaller forms are first shifted, and only then wedged or
fiber-producted. 
Implicit in these formula are the $\pm$ signs multiplying the canonical forms in the RHS. They
depend on orientation conventions for the boundary cells, bridge parameters,
internal little-group quotients, and pushforwards. Practically they can be determined up to a single global sign by the requirement of spurious poles' cancellations.

\subsection{Strong positivity and a semi-algebraic description}
\label{subsec:reflected-strong-positivity-sign-flips}
Mandelstam variables, and the variables $\mu_r$, need not to be of definite sign on the amplituhedron. A similar phenomenon occurs in the usual momentum amplituhedron. Galashin \cite{Galashin2024AmplituhedraOrigami} observed that by strengthening the positivity requirements on the external data one can obtain momentum amplituhedra on which Mandelstam variables have a definite sign. This special case of momentum amplituhedra has several desirable properties. First, under this strong positivity one can show that zero loci of Mandelstam invariants are topological boundaries of the amplituhedra. This assumption is crucial for proving BCFW tiling conjecture. Additionally, with this assumption one can prove the sign flip characterization of the momentum amplituhedron, conjectured in \cite{DamgaardFerroLukowskiParisi2019MomentumAmplituhedron}, following \cite{ArkaniHamedThomasTrnka2018Unwinding}. 
\cite{OrenPerlsteinTessler2025ABJMBCFWTiling} adjusted this property to the orthogonal (ABJM) amplituhedron case. We now explain the modification required for the reflected Lagrangian case.


The first consequence will be that under this condition the primary reflected boundary functionaries $\mu_r,
1\leq r\leq n-1,$ of \eqref{eq:mu} and $s_{[a,b]},$ for $
1\leq a<b\leq n,$ will have definite sign. This will be used in \cite{MS} to prove that their zero loci yield the real codimension $1$ boundaries of the Lagrangian amplituhedra, and the BCFW tiling theorem for this case. It will also imply that other Mandelstam variables to not give rise to codimension $1$ boundaries, and will allow us to formulate the sign flip description of $\Lampl^\refl_n.$

\subsubsection{Strong positivity}

The positivity input comes from expanding the reflected boundary functionaries in
a positive basis on the domain Grassmannian.  Let
\[
F(C;W)
\]
denote either a half-contained Mandelstam \(s_{[a,b]}\) or a symmetric square
\(s_{J_r}=\mu_r^2\).  Using Cauchy--Binet, the spinor brackets appearing in
\(F\) can be expanded into products of Pl\"ucker coordinates of \(C\) and
Pl\"ucker coordinates of \(W\).  After collecting the \(C\)-dependent terms in
the Temperley--Lieb immanant basis \cite{skandera2004inequalities,rhoades2005temperley,lam2015dimers}, one obtains an expansion of the form
\begin{equation}
\label{eq:reflected-TL-expansion}
F(C;W)
=
\sum_{\alpha}
c^F_{\alpha}(W)\,\Delta_{\alpha}(C).
\end{equation}
Here \(\alpha\) indexes the relevant Temperley--Lieb, or planar
noncrossing, data.  The coefficient \(c^F_{\alpha}(W)\) depends only on the
external datum \(W\).  The factor \(\Delta_{\alpha}(C)\) is a
Temperley--Lieb immanant: it is a specific polynomial in the maximal
Pl\"ucker coordinates of \(C\), not usually a single Pl\"ucker coordinate.  The
key fact is that
\[
\Delta_{\alpha}(C)\geq0
\qquad
\text{for all } C\in\Gr_{\geq}(n,2n).
\]
We recall the Cauchy--Binet step and the notion of immanants in
\cref{app:cauchy-binet-TL-immanants} of the appendix.

\begin{definition}[Strong reflected positivity]
\label{def:reflected-strong-positivity}
A reflected external datum \(W\in\Mat(2n,n+2)\) is called \emph{strongly positive} if
\(W\) is positive in the ordinary sense and, for every primary reflected
boundary functionary \(F\), every nonzero coefficient in
\eqref{eq:reflected-TL-expansion} is strictly positive:
\begin{equation}
\label{eq:strong-positive-coefficients}
c^F_{\alpha}(W)>0
\qquad
\text{whenever } c^F_{\alpha}(W)\not\equiv0.
\end{equation}
\end{definition}

It follows immediately from \eqref{eq:reflected-TL-expansion} and the
nonnegativity of the immanants that, for strongly positive \(W\),
\[
s_{[a,b]}(C;W)\geq0
\]
and
\[
s_{J_r}(C;W)=\mu_r(C;W)^2\geq0
\]
for all
\[
C\in\LG_{\geq}^{\refl}(n,2n).
\]
On the interior, these inequalities are strict unless the corresponding
boundary functionary is forced to vanish.

The immanant expansion controls the square \(s_{J_r}\).  To obtain a
semi-algebraic description using the reduced functions \(\mu_r\), we also need
to fix their sign. Since, under strong positivity assumptions, \(s_{J_r}=\mu_r^2\) is positive in the
interior of $\Lampl_n^\refl$, \(\mu_r\) cannot vanish in the interior, if $W$ is strongly
positive. To determine its sign one needs to compute in a single amplituhedron point in which it is non zero. It is thus easy to verify that, for strongly positive $W,$ 
\begin{equation}
\label{eq:reflected-mu-sign}
(-1)^n\mu_r\geq0,
\qquad
1\leq r\leq n-1.
\end{equation}
On the interior,
\[
(-1)^n\mu_r>0.
\]

\subsubsection{A construction of strongly positive external data}
\label{subsubsec:reflected-strong-positive-construction}

We now explain the construction of the reflected external data used in the
computations.  Start with a compatible pair
\[
\mathcal U\subset\mathbb R^{2n},
\qquad
\mathcal W\subset\mathbb R^{2n},
\]
where
\[
\dim U=n-2,
\qquad
\dim\mathcal W=n+2,
\qquad
\mathcal W=\mathcal U^{\perp_{E_{\refl}}}.
\]
The reflected external matrix
\[
W\in\Mat(2n,n+2)
\]
is any matrix whose columns span \(\mathcal W\).

A sparse compatible seed is obtained as follows.  For
\[
1\leq i\leq n-2
\]
set
\[
f_i=e_i+(-1)^{n-i}e_{\bar i}.
\]
Define
\[
\mathcal{U}_0=\operatorname{span}(f_1,\ldots,f_{n-2})
\]
and
\[
\mathcal W_0=\mathcal{U}_0^{\perp_{E_{\refl}}}.
\]
Equivalently,
\[
\mathcal W_0
=
\operatorname{span}
\bigl(
f_1,\ldots,f_{n-2},
e_{n-1},e_n,e_{n+1},e_{n+2}
\bigr).
\]
Thus
\[
\dim \mathcal{U}_0=n-2,
\qquad
\dim\mathcal W_0=n+2,
\qquad
\mathcal W_0=\mathcal{U}_0^{\perp_{E_{\refl}}}.
\]
Let
\[
M=M_{\ell}\cdots M_1
\]
be a long-enough word in reflected positivity-preserving symplectic bridge matrices, with
all bridge parameters positive.  We set
\[
\mathcal{U}(M):=M \mathcal{U}_0,
\qquad
\mathcal W(M):=M\mathcal W_0.
\]
Since every generator is symplectic,
\[
M E_{\refl}M^{\mathsf T}=E_{\refl},
\]
we still have
\[
\mathcal W(M)=\mathcal{U}(M)^{\perp_{E_{\refl}}}.
\]
Choosing \(W(M)\) with column span \(\mathcal W(M)\), we obtain a folded
reflected external datum.  
amplituhedron map through
\[
C(MW_0)=(CM)W_0.
\]


The sparse seed lies on the boundary of the strongly positive region.  The role of the
positive word is to move it into the interior while preserving the folded
compatibility relation. Similar to \cite[Section 6.2]{Galashin2024AmplituhedraOrigami} and \cite[Section 7.3]{OrenPerlsteinTessler2025ABJMBCFWTiling} there are words $M$ such that 
\[\mathcal U(M)\in\Gr_{>}(n-2,2n),\qquad\mathcal W(M)\in\Gr_{>}(n+2,2n),
\]
and all nonzero coefficients in \eqref{eq:reflected-TL-expansion} are
strictly positive.  
exact computations.

\subsubsection{Consequences for boundary functionaries}

Strong positivity implies \eqref{eq:reflected-mu-sign} and the nonnegativity of all Mandelstam variables.  This
also explains why asymmetric axis-crossing Mandelstams do not give additional
facets: Recall the identity \eqref{eq:asymmetric-Mandelstam-identity-strong-section}:
\begin{equation*}
s_{[\bar b,a]_{\cyc}}
=
s_{[a+1,b]}+\mu_a\mu_b
\qquad
(a+1<b).
\end{equation*}
Because all \(\mu_r\)'s have the same fixed sign, the product
\[
\mu_a\mu_b
\]
is nonnegative.  Hence, under strong positivity, the vanishing of
\[
s_{[\bar b,a]_{\cyc}}
\]
forces simultaneous vanishing of more than one primary boundary functionaries.
Thus these asymmetric Mandelstams cut out higher-codimension intersections,
not new codimension-one supports. 

\subsubsection{Twisted-cyclic sign flips}
We now state the sign-flip part of the intrinsic kinematic description.  For a cyclic
anchor \(i\), define the twisted square-bracket sequence
\[
\sigma_i^{[\,]}(\widetilde\lambda)
=
\bigl(
[i,\,i{+}1]_{\mathrm{tw}},
[i,\,i{+}2]_{\mathrm{tw}},
\ldots,
[i,\,i{+}2n{-}1]_{\mathrm{tw}}
\bigr),
\]
where all labels are read cyclically modulo \(2n\).  The subscript
\(\mathrm{tw}\) means that, when the cyclic order crosses the seam
\(2n\mid1\), the bracket is multiplied by the usual cyclic sign $
(-1)^{n-1}.$ 
Similarly, define
\[
\sigma_i^{\langle\,\rangle}(\lambda)
=
\bigl(
\langle i,\,i{+}1\rangle_{\mathrm{tw}},
\langle i,\,i{+}2\rangle_{\mathrm{tw}},
\ldots,
\langle i,\,i{+}2n{-}1\rangle_{\mathrm{tw}}
\bigr).
\]
The same argument as the one used in \cite{ArkaniHamedThomasTrnka2018Unwinding,KarpWilliams2019M1Amplituhedron} shows that 
\begin{equation}
\label{eq:reflected-square-sign-flips}
\overline{\operatorname{var}}
\bigl(
\sigma_i^{[\,]}(\widetilde\lambda)
\bigr)
=
n
\end{equation}
for every cyclic anchor \(i\).  Here, following \cite{KarpWilliams2019M1Amplituhedron}, \(\overline{\operatorname{var}}\) means the number of
sign changes after replacing zeros by non zero elements, in a way which maximizes the number of sign changes in the sequence.  Equivalently, using the folded relation
\[\langle ij\rangle=(-1)^{i+j}[\bar i\,\bar j],\]
one has
\begin{equation}
\label{eq:reflected-angle-sign-flips}
\overline{\operatorname{var}}
\bigl(
\mathcal \sigma_i^{\langle\,\rangle}(\lambda)
\bigr)
=
n-2.
\end{equation}
\begin{remark}
The condition is genuinely cyclic: no fixed sign-flip count was found for the
two half-sequences obtained by cutting the labels at the reflection axis.
\end{remark}

\subsubsection{The semi-algebraic description at strongly positive $W$}
We can now describe the reflected Lagrangian amplituhedron $\Lampl^\refl_n(W),$ for strongly positive $W$, in the kinematic space/$B-$amplituhedron picture. This description will be proven in \cite{MS}.

Let $W$ be a strongly positive matrix. The $B-$reflected Lagrangian amplituhedron is the collection of all \[\tilde\lambda=(\tilde\lambda_i)_{i=1,\ldots,2n}\in\Gr_{2,\operatorname{colspan}(W)}\]such that the following hold
\begin{itemize}
\item For $\lambda_i=(-1)^i\widetilde\lambda_{\bar i}$ \[
\sum_i \lambda_i\widetilde\lambda_i^{\mathsf T}=0.
\]
\item \begin{equation}
\label{eq:reflected-semialgebraic-inequalities}
(-1)^n\mu_r>0,
\qquad
s_{[a,b]}>0,
\end{equation}
\item The (cyclic) sign flip condition
\eqref{eq:reflected-square-sign-flips} for every cyclic anchor \(i\).  
\end{itemize}
 
\subsection{Examples}
\subsubsection{The two tiles of \(\Lampl_3^{\refl}(W)\)}
\label{subsec:reflected-n3-example}

We finish the reflected section with the first nontrivial example.  Here
\(n=3\), so the domain is
\[
\LG_{\geq}^{\refl}(3,6),
\qquad
\dim \LG^{\refl}(3,6)=6,
\]
while the reflected Lagrangian amplituhedron has dimension
\[
\dim \Lampl_3^{\refl}(W)=5.
\]
The reflection is
\[
\bar1=6,
\qquad
\bar2=5,
\qquad
\bar3=4.
\]
The basic reflected boundary functionaries are
\[
\mu_1=[16],
\qquad
\mu_2=[16]-[25]=-[34],
\]
and
\[
s_{12}=[12][56],
\qquad
s_{23}=[23][45],
\]
\[
s_{123}
=
[12][56]-[13][46]+[23][45].
\]
The two-particle Mandelstams are reducible, whereas \(s_{123}\) is the
three-particle half-interval Mandelstam.

For strongly positive external data, the reflected BCFW recursion for
\(\Lampl_3^{\refl}(W)\) has two $5-$dimensional domain cells.  We denote
them by
\(
\mathcal C_{\mathrm I}^{\refl},
\qquad
\mathcal C_{\mathrm{II}}^{\refl}.
\)
Their images tile \(\Lampl_3^{\refl}(W)\):
\[
\Lampl_3^{\refl}(W)
=
\overline{\Phi_W(\mathcal C_{\mathrm I}^{\refl})
\sqcup
\Phi_W(\mathcal C_{\mathrm{II}}^{\refl})},
\]where $\overline{X}$ is the closure of $X$.
\subsubsection{The first cell}
For
\(
\alpha,\beta,\gamma,x,t>0,
\)
define
\[
C_{\mathrm I}(\alpha,\beta,\gamma;x,t)
=
\begin{pmatrix}
1&x&0&0&0&t\\
0&1&0&-\alpha\beta&-\alpha&-\alpha x\\
0&0&1&\alpha\beta^2+\gamma&\alpha\beta&\alpha\beta x
\end{pmatrix}.
\]
A direct calculation gives
\[
C_{\mathrm I}E_{\refl}C_{\mathrm I}^{\mathsf T}=0.
\]
Its only identically vanishing ordered maximal minors are
\[
\Delta_{156}(C_{\mathrm I})=0,
\qquad
\Delta_{345}(C_{\mathrm I})=0.
\]
These two equations form one Lagrangian pair under
\(
I\longmapsto\overline{I^c},
\)
so the cell is five-dimensional.

The two parameters
\(
t,~
x
\)
are the parameters which touch physical Mandelstam-type boundaries.  More
precisely:
\(
t=0
\)
is the axis-bridge boundary. 
It is the branch 
\[
[12]=0
\]
of the reducible Mandelstam
\[
s_{12}=[12][56].
\]
Thus \(t\) is the BCFW parameter for the affected component of the
\(s_{12}=0\) boundary.

On the other hand,
\(
x=0
\)
is the paired-bridge boundary corresponding to
\(
\mu_1=0.
\)
Thus \(x\) is the BCFW parameter for the reduced symmetric boundary.

The remaining parameters
\(
\alpha,\beta,\gamma
\)
are coordinates which do not represent
top-level Mandelstam variable.  

Up to an overall orientation sign, the logarithmic form on the domain cell is
\[
\omega_{\mathrm I}
=
d\log t\wedge d\log\alpha\wedge d\log\beta
\wedge d\log\gamma\wedge d\log x.
\]

\subsubsection{The second cell}

For
\[
A,B,G,u,z>0,
\]
define
\[
C_{\mathrm{II}}(A,B,G;u,z)
=
\begin{pmatrix}
1&z&0&-AB&0&A\\
0&u&1&AB^2+G&0&-AB\\
0&0&0&u&1&z
\end{pmatrix}.
\]
Again,
\[
C_{\mathrm{II}}E_{\refl}C_{\mathrm{II}}^{\mathsf T}=0.
\]
Its only identically vanishing ordered maximal minor is
\[
\Delta_{123}(C_{\mathrm{II}})=0.
\]
This minor is self-dual under
\(
I\longmapsto\overline{I^c},
\)
so this is again a $5-$dimensional reflected Lagrangian cell.

The parameter
\(
z
\)
is a physical BCFW parameter.  At
\(
z=0,
\)
the cell lies on the affected branch
\[
[45]=0
\]
of
\[
s_{23}=[23][45].
\]
The other branch,
\(
[23]=0,
\)
is invariant under this bridge and does not give a BCFW contribution for this
choice of shift.

The second physical BCFW parameter is less visible in the coordinates
\((A,B,G)\).  Set
\[
P=AB,
\qquad
H=AB^2+G,
\qquad
\tau=\frac{AG}{AB^2+G}=\frac{AG}{H}.
\]
Equivalently,
\[
A=\frac{P^2+\tau H}{H},
\qquad
B=\frac{PH}{P^2+\tau H},
\qquad
G=\frac{\tau H^2}{P^2+\tau H}.
\]
The logarithmic Jacobian is
\[
d\log A\wedge d\log B\wedge d\log G
=
d\log\tau\wedge d\log P\wedge d\log H.
\]
In these variables,
\[
C_{\mathrm{II}}
=
\begin{pmatrix}
1&z&0&-P&0&P^2/H+\tau\\
0&u&1&H&0&-P\\
0&0&0&u&1&z
\end{pmatrix}.
\]
Then
\[
\tau=0
\]
is the generic component of the three-particle Mandelstam boundary
\(
s_{123}=0.
\)
Thus \(\tau\) is the axis-bridge BCFW parameter for the \(s_{123}\) boundary.

The remaining parameters $
P,H,u
$
are coordinates whose zero loci are not codimension $1$ boundaries of the tile.

Up to orientation, the logarithmic form may be written as
\[
\omega_{\mathrm{II}}
=
d\log z\wedge d\log\tau\wedge d\log P
\wedge d\log H\wedge d\log u.
\]
Equivalently, in the original coordinates,
\[
\omega_{\mathrm{II}}
=
d\log z\wedge d\log A\wedge d\log B
\wedge d\log G\wedge d\log u.
\]

\subsubsection{The two BCFW presentations}

The same two domain cells appear from two different choices of reflected BCFW
bridge.
For the outer axis bridge, the physical boundary parameters are
\(
t\) and \(
\tau:
\)
\[
\mathcal C_{\mathrm I}^{\refl}
\quad\text{comes from}\quad
s_{12}=0,
\]
with \(t\) the bridge parameter, while
\[
\mathcal C_{\mathrm{II}}^{\refl}
\quad\text{comes from}\quad
s_{123}=0,
\]
with \(\tau\) the bridge parameter.

For the paired bridge at the cut \(1\mid2\), the physical boundary parameters
are
$x$ and $z:$
\[
\mathcal C_{\mathrm I}^{\refl}
\quad\text{comes from}\quad
\mu_1=0,
\]
with \(x\) the bridge parameter, while
\[
\mathcal C_{\mathrm{II}}^{\refl}
\quad\text{comes from}\quad
s_{23}=0,
\]
with \(z\) the bridge parameter.

Thus the bridge choice changes which boundary face is used to generate each
tile, but it does not change the two tiles themselves. This phenomenon is special to $n=3.$ The bookkeeping is:
\[
\begin{array}{c|c|c|c}
\text{cell}
&
\text{parameters touching physical boundaries}
&
\text{physical boundaries}
&
\text{remaining parameters}
\\ \hline
\mathcal C_{\mathrm I}^{\refl}
&
t,\ x
&
s_{12},\ \mu_1
&
\alpha,\beta,\gamma
\\[1mm]
\mathcal C_{\mathrm{II}}^{\refl}
&
z,\ \tau
&
s_{23},\ s_{123}
&
P,H,u
\end{array}
\]

\subsubsection{Boundary study at \(n=4\)}
\label{subsec:reflected-n4-boundary-check}
We also record the first rank in which the reflected boundary bookkeeping
already shows the two main phenomena: symmetric square boundaries and
asymmetric Mandelstams which do not give new codimension-one supports.

For \(n=4\) the labels are paired by
\[
1\leftrightarrow 8,
\qquad
2\leftrightarrow 7,
\qquad
3\leftrightarrow 6,
\qquad
4\leftrightarrow 5.
\]
The scalar folded momentum-conservation equation is
\[
[18]-[27]+[36]-[45]=0.
\]
Thus
\[
\mu_1=[18],
\]
\[
\mu_2=[18]-[27],
\]
and
\[
\mu_3=[18]-[27]+[36]=[45],
\qquad
\mu_4=0.
\]
The corresponding symmetric intervals are
\[
J_1=[8,1]_{\cyc},
\qquad
J_2=[7,2]_{\cyc},
\qquad
J_3=[6,3]_{\cyc},
\]
and
\[
s_{J_r}=\mu_r^2.
\]
The half-contained two-particle Mandelstams are:
\[
s_{[1,2]}=[12][78],
\]
\[
s_{[2,3]}=[23][67],
\]
\[
s_{[3,4]}=[34][56].
\]
They are reducible into collinear components.  The first genuinely
higher-particle half-contained examples are
\[
s_{[1,3]}
=
[12][78]-[13][68]+[23][67],
\]
\[
s_{[2,4]}
=
[23][67]-[24][57]+[34][56],
\]
and
\[
\begin{aligned}
s_{[1,4]}
={}&
[12][78]-[13][68]+[14][58]
\\
&\quad
+[23][67]-[24][57]+[34][56].
\end{aligned}
\]
Now consider an asymmetric interval crossing a reflection axis.  \eqref{eq:asymmetric-Mandelstam-identity-strong-section} gives\[
s_{[6,1]_{\cyc}}
=
s_{[2,3]}+\mu_1\mu_3.
\]
Under the strong sign assumptions,
\[
s_{[2,3]}\geq0,
\qquad
\mu_1\geq0,
\qquad
\mu_3\geq0.
\]
Therefore the vanishing of \(s_{[6,1]_{\cyc}}\) forces simultaneous vanishing
of already existing primary boundary functionaries.  Thus this asymmetric
Mandelstam does not define a new codimension-one support.

The domain dimensions also match the general boundary-cell formulas.  The full
domain has
\[
\dim \LG^{\refl}(4,8)=10.
\]
For the central symmetric boundary
\(
\mu_2=0,
\)
the maximal block cell is
\[
\LG_{\geq}^{\refl}(2,4)
\times
\LG_{\geq}^{\refl}(2,4),
\]
of dimension
\[
3+3=6,
\]
hence domain codimension \(4\).  For the two other symmetric boundaries,
\(\mu_1=0\) and \(\mu_3=0\), the block dimensions are
\[
\dim \LG^{\refl}(1,2)+\dim \LG^{\refl}(3,6)
=
1+6=7,
\]
again mapping to codimension-one boundary supports in the amplituhedron.

For the half interval
\[
I=[1,3],
\qquad
m=3,
\qquad
h=n-m+1=2,
\]
the source construction has three sectors
\[
(r,s)=(0,2),(1,1),(2,0).
\]
Only the middle sector \((1,1)\) is a nondegenerate codimension-one
factorization sector.  The endpoint sectors \(r=0\) and \(s=0\) force one of
the ordinary momentum factors to have rank \(1\), respectively corank \(1\),
and hence impose extra collinear equations.  Their domain dimensions are
\[
5,\quad 6,\quad 5,
\]
so only the middle sector has the expected boundary dimension \(2n-2=6\).

For the longest half interval
\[
I=[1,4],
\qquad
m=4,
\qquad
h=1,
\]
there are four maximal cells, with
\(
r=0,1,2,3,
~
s=3-r.
\)
The ordinary factors are
\(
\Gr_{\geq}(r+1,5),
\)
and the smaller reflected core is
\(
\LG_{\geq}^{\refl}(1,2).
\)
The domain dimensions are
\(
4, ~6,~ 6,~ 4.
\), respectively. Again the sectors $r=0,r=3$ do not map to a codimension $1$ boundary stratum.

\appendix
\section{Cauchy--Binet expansions and Temperley--Lieb immanants}
\label{app:cauchy-binet-TL-immanants}

In this appendix we provide the algebraic background needed for in
\cref{subsec:reflected-strong-positivity-sign-flips}.  The purpose is not to
introduce the full theory of Temperley--Lieb immanants
\cite{rhoades2005temperley,skandera2004inequalities,lam2015dimers}, but only
to clarify the expansion
\eqref{eq:reflected-TL-expansion}.  We follow
\cite{Galashin2024AmplituhedraOrigami}, with the folded modifications needed
for the reflected Lagrangian setting.

\subsection{Cauchy--Binet for the projected brackets}

Let
\[
C\in\Gr(n,2n),
\qquad
W\in\Mat(2n,n+2),
\qquad
Y=CW.
\]
For \(a,b\in[2n]\), the square bracket
\([ab]=\det(\widetilde\lambda_a,\widetilde\lambda_b)\) can be written, up to a
common normalization, as
\begin{equation}
\label{eq:appendix-bracket-YW}
[ab]
=
\det
\begin{pmatrix}
Y\\
W_a\\
W_b
\end{pmatrix}.
\end{equation}
Here \(W_a,W_b\) are the corresponding rows of \(W\).  Since
\[
Y=CW,
\]
we may rewrite the matrix in \eqref{eq:appendix-bracket-YW} as
\[
\begin{pmatrix}
C\\
e_a\\
e_b
\end{pmatrix}
W.
\]
Applying Cauchy--Binet gives
\begin{equation}
\label{eq:appendix-CB-general}
\det(AB)
=
\sum_{K}
\Delta_K(A)\Delta_K(B),
\end{equation}
where the sum is over subsets \(K\) of the appropriate size.  In the present
case this becomes
\begin{equation}
\label{eq:appendix-bracket-CB}
[ab]
=
\sum_{\substack{I\in\binom{[2n]}{n}\\ I\cap\{a,b\}=\varnothing}}
\epsilon(I;a,b)\,
\Delta_I(C)\,
\Delta_{I\cup\{a,b\}}(W).
\end{equation}
The sign
\[
\epsilon(I;a,b)\in\{\pm1\}
\]
is the sign obtained by reordering the rows of
\[
\begin{pmatrix}
C\\
e_a\\
e_b
\end{pmatrix}
\]
into the ordered set \(I\cup\{a,b\}\).  Thus each spinor bracket is a linear
combination of Pl\"ucker coordinates of \(C\), with coefficients given by
Pl\"ucker coordinates of the external datum \(W\).  It has a definite sign
when \(a,b\) are cyclically consecutive.

A Mandelstam variable is quadratic in spinor brackets:
\begin{equation}
\label{eq:appendix-Mandelstam-spinor}
s_I
=
\sum_{\substack{a<b\\ a,b\in I}}
\langle ab\rangle[ab].
\end{equation}
In the reflected Lagrangian model, the angle brackets are folded square brackets:
\begin{equation}
\label{eq:appendix-folded-angle-square-refl}
\langle ab\rangle=(-1)^{a+b}[\bar a\,\bar b].
\end{equation}
Substituting \eqref{eq:appendix-bracket-CB} into
\eqref{eq:appendix-Mandelstam-spinor}, and then using
\eqref{eq:appendix-folded-angle-square-refl}, one obtains an expression of the
form
\begin{equation}
\label{eq:appendix-quadratic-plucker-C}
s_I(C;W)
=
\sum_{A,B}
d^I_{A,B}(W)\,\Delta_A(C)\Delta_B(C).
\end{equation}
Here the coefficients \(d^I_{A,B}(W)\) are explicit quadratic expressions in
the Pl\"ucker coordinates of \(W\).  Formula
\eqref{eq:appendix-quadratic-plucker-C} is the Cauchy--Binet origin of the
external coefficients which appear in the main text.

\subsection{Temperley--Lieb immanants}

Products
\[
\Delta_A(C)\Delta_B(C)
\]
are not the most useful way to read positivity.  The positive basis adapted to
planar Mandelstam functions is the Temperley--Lieb immanant basis.

For our purposes, a Temperley--Lieb immanant is a polynomial
\(
\Delta_{\alpha}(C)
\)
in the Pl\"ucker coordinates of \(C\), indexed by a Temperley--Lieb diagram
\(\alpha\).  One concrete way to define these functions is through planar
networks.  Let \(N\) be a reduced planar network whose boundary measurement is
the point \(C\in\Gr_{\geq}(n,2n)\).  A Temperley--Lieb diagram \(\alpha\)
specifies a planar connectivity pattern among boundary labels.  The
corresponding immanant is obtained by summing the weights of all collections of
paths in \(N\) whose boundary connectivity is \(\alpha\):
\begin{equation}
\label{eq:appendix-TL-path-definition}
\Delta_{\alpha}(C_N)
=
\sum_{\mathcal P:\,\operatorname{conn}(\mathcal P)=\alpha}
\operatorname{wt}(\mathcal P).
\end{equation}
Here \(\operatorname{wt}(\mathcal P)\) is the product of the edge weights
appearing in the path collection.  This expression is independent of the
chosen planar network chart and therefore defines a regular function on the
Grassmannian.  Since all weights in a positive network are nonnegative,
\eqref{eq:appendix-TL-path-definition} makes the total nonnegativity of
\(\Delta_\alpha\) transparent, which is crucial for amplituhedron applications.

The relation with Pl\"ucker coordinates is as follows.  Ordinary Pl\"ucker
coordinates are the special case in which the path collection has a fixed set
of sources and sinks.  Products of Pl\"ucker coordinates correspond to pairs
of such path collections.  When the induced connectivity pattern has crossings,
one resolves it using the Temperley--Lieb, or planar, straightening relations.
Equivalently, for every product of maximal Pl\"ucker coordinates one has a
finite expansion
\[
\Delta_A(C)\Delta_B(C)
=
\sum_{\alpha}
m_{A,B}^{\alpha}\,\Delta_{\alpha}(C),
\]
with coefficients \(m_{A,B}^{\alpha}\) independent of \(C\) and \(W\).  Thus
the passage from the Cauchy--Binet expression
\eqref{eq:appendix-quadratic-plucker-C} to the immanant expansion
\begin{equation}
\label{eq:appendix-TL-rewrite}
s_I(C;W)
=
\sum_{\alpha}
c^I_{\alpha}(W)\,\Delta_{\alpha}(C)
\end{equation}
is simply this fixed change of basis on the source side.  The external
coefficients \(c^I_\alpha(W)\) are obtained by applying the same linear change
of basis to the Cauchy--Binet coefficients \(d^I_{A,B}(W)\).
The same discussion applies to the symmetric square functions in the reflected
model,
\(
s_{J_r}=\mu_r^2.
\)
Thus the functions used in the reflected strong-positivity section have
expansions
\begin{equation}
\label{eq:appendix-reflected-unified-TL-expansion}
F(C;W)
=
\sum_{\alpha}
c^F_{\alpha}(W)\,\Delta_{\alpha}(C),
\end{equation}
where
\[
F\in
\{\,s_{[a,b]}:1\leq a<b\leq n\,\}
\cup
\{\,s_{J_r}:1\leq r\leq n-1\,\}.
\]

\subsection{Strong positivity revisited}

With the notation of
\eqref{eq:appendix-reflected-unified-TL-expansion}, strong reflected
positivity means that
\[
c^F_{\alpha}(W)>0
\]
for every nonzero term that appears in the expansion of every reflected
primary function \(F\).  Combining this with
\eqref{eq:appendix-reflected-unified-TL-expansion} and the nonnegativity of
Temperley--Lieb immanants gives
\[
F(C;W)\geq0
\qquad
\text{for all }C\in\LG_{\geq}^{\refl}(n,2n).
\]
This is the positivity input used in
\cref{subsec:reflected-strong-positivity-sign-flips}.

\section{Computational evidence}
\label{app:computational-evidence}

We supported our theoretical results with computational evidence.  Below is a
concise summary of the experiments relevant to the reflected Lagrangian
amplituhedron.

\paragraph{BCFW tilings in the reflected model.}
We tested experimentally, up to \(n=5\), that the reflected Lagrangian
amplituhedron admits BCFW tilings, both for positive and for strongly positive
external data.  The experimental method was to randomly choose many points in
the big cells of the Lagrangian Grassmannian, map them to the amplituhedron,
and count preimages inside the domain cells belonging to a given fully expanded
BCFW collection.  We used several choices of external data and several BCFW
collections.  In all experiments, for each BCFW collection, every tested point
had a unique preimage in a unique cell of that collection.  To this end we
wrote an algorithm for finding preimages, based on the promotion methods of
\cite{EvenZoharLakrecTessler2025BCFWTriangulation}.

\paragraph{Boundary cancellation.}
Up to \(n=6\), we tested all boundaries of certain collections of BCFW cells.
For the collections we checked, the codimension-one domain boundaries were of
three types: boundaries mapping to the zero locus of a primary reflected
boundary function; boundaries cancelling in pairs as spurious poles; and
boundaries mapping to lower dimension under the amplituhedron map.

\paragraph{Sign flips.}
For every tested interior point of ranks
\(n=3,4,5,6\) and every cyclic anchor, the twisted-cyclic square-bracket
sequence had \(n\) sign flips, while the folded angle-bracket sequence had
\(n-2\) sign flips.  Here we use the conventions of the paper, in which
\(\widetilde\lambda=C^\perp\cap\mathcal W\) is the \(B\)-plane.  These
experiments did not assume strong positivity of external data; this direction
of the sign-flip property follows from the original arguments of
\cite{ArkaniHamedThomasTrnka2018Unwinding}.  No fixed sign-flip count was
found for the two half-sequences obtained by cutting the labels at the
reflection axis.

\paragraph{Canonical forms.}
As a sanity check for our implementation of canonical forms, we first tested it
on the ordinary momentum amplituhedron.  We pushed the canonical \(d\log\)
forms of the BCFW cells to a common target frame of  and compared distinct shifted
triangulations at exact rational targets.  The two triangulations of
\(\mathcal M_{6,3}\) agreed at all tested points, as did shifted collections
for \(\mathcal M_{7,3}\), \(\mathcal M_{8,3}\), and \(\mathcal M_{8,4}\).
We also constructed the recursive domain amalgamation explicitly, identified
the internal gluing torus, and verified that the recursively induced domain
form agrees with the canonical permutation-bridge form up to orientation.

For the Lagrangian case, the rank-three and rank-four canonical-form sums were explicitly calculated
and were seen to be independent of the chosen recursive presentation or paired bridge. 

\bibliographystyle{alpha}
\bibliography{Lagrangian_bibli}

\end{document}